\documentclass[usenatbib]{mnras}
\usepackage{natbib,amssymb,amsmath,times,graphicx,url,aas_macros}
\usepackage{setspace}      
\usepackage{epstopdf}       
\usepackage{verbatim}       
\usepackage{bm}		  
\usepackage{fleqn}		  
\usepackage{Wurster_macros}  
\defcitealias{WursterBatePrice2018ff}{Paper I}

\newcommand{\mf}{$N=10^5$}

\newcommand{\msx}{$N=10^6$}
\newcommand{\mtsx}{$N=3\times10^6$}

\newcommand{\mtsv}{$N=3\times10^7$}

\newcommand{\mtfx}{$N\le3\times10^5$}
\newcommand{\msxx}{$N\le10^6$}
\newcommand{\mtsxx}{$N\le3\times10^6$}

\newcommand{\mtfn}{$N\ge3\times10^5$}
\newcommand{\msxn}{$N\ge10^6$}
\newcommand{\mtsxn}{$N\ge3\times10^6$}
\newcommand{\msvn}{$N\ge10^7$}

\title[On the origin of magnetic fields in stars II]{On the origin of magnetic fields in stars II: The effect of numerical resolution}

\author[Wurster, Bate, Price \& Bonnell]{James Wurster$^{1,2}$\thanks{jhw5@st-andrews.ac.uk}, Matthew R. Bate$^{2}$\thanks{mbate@astro.ex.ac.uk}, Daniel J. Price$^{3}$, and Ian A. Bonnell$^{1}$ \\
$^{1}$Scottish Universities Physics Alliance (SUPA), School of Physics and Astronomy, University of St. Andrews, North Haugh, St Andrews, Fife KY16 9SS, UK \\
$^{2}$School of Physics and Astronomy, University of Exeter, Stocker Rd, Exeter EX4 4QL, UK \\
$^{3}$Monash Centre for Astrophysics and School of Physics and Astronomy, Monash University, Vic 3800, Australia \\
}
\date{Submitted: Revised: Accepted: }
\pagerange{\pageref{firstpage}--\pageref{lastpage}} \pubyear{2021}

\begin{document}
\label{firstpage}
\bibliographystyle{mnras}
\maketitle

\begin{abstract}
Are the kG-strength magnetic fields observed in young stars a fossil field left over from their formation or are they generated by a dynamo?  Our previous numerical study concluded that magnetic fields must originate by a dynamo process.  Here, we continue that investigation by performing even higher numerical resolution calculations of the gravitational collapse of a 1~M$_\odot$ rotating, magnetised molecular cloud core through the first and second collapse phases until stellar densities are reached.  Each model includes Ohmic resistivity, ambipolar diffusion, and the Hall effect.  We test six numerical resolutions, using between $10^5$ and $3\times10^7$ particles to model the cloud.  At all but the lowest resolutions, magnetic walls form in the outer parts of the first hydrostatic core, with the maximum magnetic field strength located within the wall rather than at the centre of the core.  At high resolution, this magnetic wall is disrupted by the Hall effect, producing a magnetic field with a spiral-shaped distribution of intensity.  As the second collapse occurs, this field is dragged inward and grows in strength, with the maximum field strength increasing with resolution.  As the second core forms, the maximum field strength exceeds 1~kG in our highest resolution simulations, and the stellar core field strength exceeds this threshold at the highest resolution.  Our resolution study suggests that kG-strength magnetic fields may be implanted in low-mass stars during their formation, and may persist over long timescales given that the diffusion timescale for the magnetic field exceeds the age of the Universe.
\end{abstract}


\section{Introduction}
\label{intro}
What is the origin of magnetic fields in low-mass stars?  Observations show strong, kG-strength surface magnetic fields on low-mass stars that weaken as they age \citepeg{YangJohnskrullValenti2005,DonatiLandstreet2009,Lavail+2017,Donati+2020,Sokal+2020}.  Since young, low-mass stars are fully convective, it is generally assumed that any birth magnetic fields are quickly diffused and replaced by dynamo-generated fields \citep{ChabrierKuker2006}.  Moreover, their long-term evolution is consistent with their magnetic fields being generated by convective dynamos since stellar rotation rates also decrease with time due to the emission of magnetised winds and outflows  \citepeg{Parker1958,Schatzman1962,WeberDavis1967,Skumanich1972,Pizzolato+2003,Wright+2011,Vidotto+2014,See+2015}.  However, given the large dispersion in the observed magnetic field strengths of young stars \citepeg{Johnskrull2007,YangJohnskrull2011}, there is speculation that the magnetic fields of low-mass stars may be dominated by primordial or `fossil' magnetic fields that are implanted during the star formation process \citep{Tayler1987,Moss2003,ToutWickramasingheFerrario2004,YangJohnskrull2011}; these studies have so far failed to find any correlation between the measured magnetic field properties and the stellar properties thought to be important for dynamo action.  Therefore, the strength and geometry of magnetic fields implanted in protostars during the star formation process remains unknown. 

The formation of protostellar cores in a magnetised medium has been the focus of many numerical studies \citepeg{MachidaInutsukaMatsumoto2006,MachidaInutsukaMatsumoto2007,Tomida+2013,TomidaOkuzumiMachida2015,Machida2014,BateTriccoPrice2014,Tsukamoto+2015oa,Vaytet+2018,WursterBatePrice2018sd,WursterBatePrice2018ff,MachidaBasu2019}.  These studies typically focused on the formation of discs and outflows, and how they are affected by ideal, resistive and/or non-ideal magnetic fields.  Additionally, several studies have commented on the magnetic field strength in the stellar cores that form.  In ideal magnetohydrodynamics (MHD) studies of the formation of an isolated star, magnetic fields of $10^4 - 10^6$~G are shown to be implanted in the stellar core at birth \citepeg{MachidaInutsukaMatsumoto2006,MachidaInutsukaMatsumoto2007,BateTriccoPrice2014,WursterBatePrice2018ff,Vaytet+2018}.  This magnetic field was much stronger than expected, and largely resulted because ideal MHD was employed.

Non-ideal MHD, however, is a more realistic prescription when modelling star formation since star forming regions are only weakly ionized \citepeg{MestelSpitzer1956,NakanoUmebayashi1986,UmebayashiNakano1990}.  When adding magnetic diffusion in the form of Ohmic resistivity and/or ambipolar diffusion, the stellar core field strength was \sm$10^2$~G at birth \citepeg{MachidaInutsukaMatsumoto2007,Vaytet+2018}, which is much smaller than in ideal MHD simulations and is below the observed kG field strengths.  The dispersive Hall effect has been shown to have a significant impact on the star forming region \citepeg{Tsukamoto+2015hall,Tsukamoto+2017,WursterPriceBate2016,WursterBatePrice2018hd,WursterBateBonnell2021}, thus it must be included when investigating the magnetic properties during star formation.  When including Ohmic resistivity, ambipolar diffusion and the Hall effect, we \citep[][herein \citetalias{WursterBatePrice2018ff}]{WursterBatePrice2018ff} found that the maximum magnetic field strength of $B_\text{max} < 1$~kG initially resided in a magnetic wall \citepeg{TassisMouschovias2005b,TassisMouschovias2007a,TassisMouschovias2007b}; similar magnetic walls were presented in, e.g., \citet{TomidaOkuzumiMachida2015} and \citet{Vaytet+2018}.  In  \citetalias{WursterBatePrice2018ff}, the location of our magnetic wall moved inwards during the second collapse phase, but the location of the maximum magnetic field strength never became coincident with the centre of the stellar core.  For the duration of the simulations, the central magnetic field strength within the stellar core was $B_\text{cen} < 300$~G.  Only slightly higher field strengths were obtained when using a cosmic ray ionisation rate ten times higher than the canonical value of \zetaeq{-17}  \citep{SpitzerTomasko1968} to calculate the non-ideal MHD effects.  Artificial resistivity also affected the magnetic field strength, however, the fiducial algorithm used \citep{Price+2018phantom} yielded stronger field strengths than its more resistive predecessor \citep{TriccoPrice2013}.   Finally, \citetalias{WursterBatePrice2018ff} modelled two resolutions that differed by a factor of ten in mass resolution; decreasing the resolution decreased the maximum field strength but yielded a similar central field strength.  Therefore, in \citetalias{WursterBatePrice2018ff}, we concluded that the origin of magnetic fields in low-mass stars was not a fossil field, but generated at a later time by a dynamo action.  

Numerical resolution plays a crucial role in astrophysical simulations.  Ideally, every numerical study would include a convergence study, where simulations of higher and higher resolution are run until it becomes clear that increasing the resolution no longer affects the physical results; it would then only be the highest resolution simulations that would be analysed.  While numerical studies typically use sufficient resolution to meet the required resolution criteria \citepeg{BateBurkert1997,Truelove+1997,Nelson2006,Commercon+2008} of the object they are trying to resolve, most exclude resolution studies, typically due to limited computational resources.  Many studies, however, perform their simulations at various resolutions to understand how resolution affects their results \citepeg{Schmidt+2010,Federrath2015,WursterPriceBate2016,WursterBatePrice2018ff,WursterBatePrice2019,Cunningham+2018,Grudic+2021}, while several have shown that their results remain unconverged, even when using the highest resolution feasible \citepeg{MacLow+1998,MeruBate2011converge,Joos+2013,BateTriccoPrice2014,Hosokawa+2016,HaugbollePadoanNordlund2018,Meyer+2018,Hennebelle2018}, and a few have shown that convergence has been reached \citepeg{Commercon+2008,MeruBate2012,LeeHennebelle2018a,LeeHennebelle2018b,YamamotoOkamotoSaitoh2021}\footnote{The preceding list contains examples from cluster formation, star formation, disc and turbulence studies.  Neither the list nor the topics is exhaustive.}.  Although convergence studies are required to produce robust conclusions, this is not always possible.

Increasing numerical resolution tends to increase the magnetic field strength in a newly formed protostar  \citepeg{BateTriccoPrice2014,WursterBatePrice2018ff} since the protostar is better resolved and its properties (including gas density and magnetic field strength) suffer less smoothing.  Therefore, the question arises as to whether or not the conclusions of \citetalias{WursterBatePrice2018ff} will hold at even higher resolutions.  The magnetic field strength should be less dependent on resolution when employing non-ideal MHD since the non-ideal processes represent physical dissipation that are not directly related to resolution.  Artificial resistivity should also become less important for increasing resolution since it is a second-order numerical term, and is generally weaker than artificial dissipation \citepeg{WursterPriceBate2016}.  However, even when including non-ideal MHD, there were resolution effects at the resolutions previously investigated \citepeg{WursterBatePrice2018ff,WursterBate2019res}.

In this paper, we investigate the effect of resolution and build upon the analysis of \citetalias{WursterBatePrice2018ff}.  We present six models with mass resolutions varying by a factor of 300 between our highest and lowest resolution model; our highest resolution model has a mass resolution 10 times higher than the fiducial simulation presented in \citetalias{WursterBatePrice2018ff}.  In Section~\ref{sec:methods}, we summarise our methods and in Section~\ref{sec:ic} we present our initial conditions.  We present our results in Section~\ref{sec:results}, discuss timescales and initial environments in \secref{sec:disc}, and conclude in Section~\ref{sec:conc}.

\section{Methods}
\label{sec:methods}

Our method is identical to that from \citetalias{WursterBatePrice2018ff}.  We solve the self-gravitating, radiation non-ideal magnetohydrodynamics equations using \textsc{sphNG}, which is a three-dimensional Lagrangian smoothed particle hydrodynamics (SPH) code that originated from \citet{Benz1990}.  Over the past 30 years, the code has been substantially modified to include (e.g.) a consistent treatment of variable smoothing lengths \citep{PriceMonaghan2007}, individual time-stepping \citep{BateBonnellPrice1995}, radiation as flux limited diffusion \citep{WhitehouseBateMonaghan2005,WhitehouseBate2006}, magnetic fields \citep[for a review, see][]{Price2012}, and non-ideal MHD \citep{WursterPriceAyliffe2014,WursterPriceBate2016}.  Gravitational forces are calculated using a binary tree, where the gravitational potential is softened using the SPH kernel such that the softening varies with the smoothing length \citep{PriceMonaghan2007}.  For stability of the magnetic field, we use the source-term subtraction approach \citep{BorveOmangTrulsen2001}, constrained hyperbolic/parabolic divergence cleaning \citep{TriccoPrice2012,TriccoPriceBate2016}, and artificial resistivity \citep[as described in][]{Price+2018phantom}\footnote{This artificial resistivity is generally weaker than the resistivity algorithm in \citet{TriccoPrice2013}, as compared in \citetalias{WursterBatePrice2018ff} and \citet{Wurster+2017}.}.  For a more detailed description, see \citet{WursterBatePrice2018sd}.

To self-consistently calculate the non-ideal MHD coefficients, we use Version 1.2.5 of the \textsc{Nicil} library \citep{Wurster2016}.  This includes cosmic ray ionisation of light and heavy metals, and thermal ionisation at high temperatures ($T \gtrsim 1000$~K).  There are three dust grain populations that differ only in charge, where the charges are $\pm1$ and 0.   We include the three non-ideal MHD terms that are important for star formation: Ohmic resistivity, ambipolar diffusion and the Hall effect.  Ohmic resistivity is calculated implicitly, as described in the appendix of \citet{WursterBatePrice2018sd}, and the remaining two terms are calculated explicitly.

There are minor differences between this version of \textsc{sphNG} and \textsc{Nicil} compared to the versions used in \citetalias{WursterBatePrice2018ff}.  For consistency, we recomputed the models from \citetalias{WursterBatePrice2018ff} so that all the models presented here are calculated using the same version of \textsc{sphNG} and \textsc{Nicil}.

\section{Initial conditions}
\label{sec:ic}

Our initial conditions are identical to those in \citetalias{WursterBatePrice2018ff}, which are the same as our previous studies \citepeg{BateTriccoPrice2014,WursterBatePrice2018sd,WursterBatePrice2018hd,WursterBatePrice2018ff,WursterBateBonnell2021}.  We initialise a spherical core of mass 1~M$_{\odot}$ with radius $R_\text{c} = 4\times10^{16}$~cm and a uniform density of $\rho_0 = 7.42\times 10^{-18}$~\gpercc; the core has an initial (isothermal) sound speed of $c_\text{s} = \sqrt{p/\rho}= 2.2\times 10^{4}$~cm~s$^{-1}$, and a solid body rotation about the $z$-axis of $\Omega = 1.77 \times 10^{-13}$~rad s$^{-1}$, which corresponds to a ratio of rotational-to-gravitational energy of $\beta_\text{r} \simeq 0.005$.  The core is placed in pressure equilibrium with a warm, low-density medium of edge length $4R_\text{c}$; magnetohydrodynamic forces are periodic across the boundary of this box but gravitational forces are not.

The entire domain is threaded with a uniform magnetic field that is parallel to and aligned with the rotation axis.  The initial magnetic field strength is $B_0 = 163\mu$G, which is equivalent to a mass-to-flux ratio of $\mu_0 = 5$ in units of the critical mass-to-flux ratio \citepeg{Mestel1999,MaclowKlessen2004}.  Although this strength is weaker than generally observed in molecular cloud cores \citep[for reviews, see, e.g.,][]{Crutcher1999,HeilesCrutcher2005,HullZhang2019}, this initial mass-to-flux ratio was chosen to match our previous studies and as a compromise since higher field strengths would be even more computationally expensive (see \secref{sec:ic:ps}).  For non-ideal MHD, we use the canonical cosmic ray ionisation rate of \zetaeq{-17}  \citep{SpitzerTomasko1968}.  

We intentionally choose an aligned orientation of the magnetic field and rotation vectors, as in \citetalias{WursterBatePrice2018ff}.  This orientation will yield a small, nearly axisymmetric protostellar disc that forms just prior to the stellar core phase, which results in a simpler analysis than if we had to account for the large disc with an $m=2$ instability that appears if the two vectors are initially anti-aligned \citepeg{WursterBatePrice2018hd,WursterBateBonnell2021}.  Moreover, the model with the aligned orientation is computationally less expensive, reaching \rhoxeq{-2} \sm5 faster than the anti-aligned orientation.  

Within a given simulation, all SPH particles have an equal mass, and the particles are initially placed on a cubic lattice.  

\subsection{Parameter space}
\label{sec:ic:ps}
In this study, we only investigate the effect of resolution.
To resolve the local Jeans mass throughout the collapse, we require at least $3\times 10^4$ particles in the sphere \citepeg{BateBurkert1997}, given our initial conditions and equation of state.  Here, we investigate six clouds, where the resolutions are given in Table~\ref{table:resolutions}.  
Throughout this paper, each simulation will be referred to by the number of particles in the cloud core.  
\begin{table}
\centering
\begin{tabular}{r r c}
\hline
$N_\text{cloud}$ & $N_\text{medium}$ & d$t_\text{sc, final}$ (yr)\\
\hline
   $              10^5$ & $5.2 \times 10^4$ & 40\\  
   $3 \times 10^5$ & $1.5 \times 10^5$ & 40\\  
   $              10^6$ & $4.8 \times 10^5$ & 40\\  
   $3 \times 10^6$ & $1.5 \times 10^6$ & 21 \\  
   $              10^7$ & $4.8 \times 10^6$ & 13\\  
   $3 \times 10^7$ & $1.5 \times 10^7$ &   4\\  
\hline
\end{tabular}
\caption{The number of particles in the cloud core of each simulation ($N_\text{cloud}$; first column) and the number of particles in the warm medium ($N_\text{medium}$; second column).  Each cloud core has a mass of 1~\Msun{}, thus $m_\text{particle} = $ \Msun$/N_\text{cloud}$.  In each simulation, every SPH particle has the same mass.  The third column lists the number of years the simulation was evolved after the formation of the stellar core.}
\label{table:resolutions}
\end{table}

Our lowest resolution model, \mf{}, matches the mass resolution in our cluster study \citep{WursterBatePrice2019}; the model with \msx{} matches that in our turbulence vs non-ideal MHD studies \citep{WursterLewis2020d,WursterLewis2020sc} and  the model with \mtsx{} matches our core-collapse studies \citep{WursterBatePrice2018sd,WursterBatePrice2018hd,WursterBatePrice2018ff,WursterBateBonnell2021}.

\begin{figure} 
\centering
\includegraphics[width=\columnwidth]{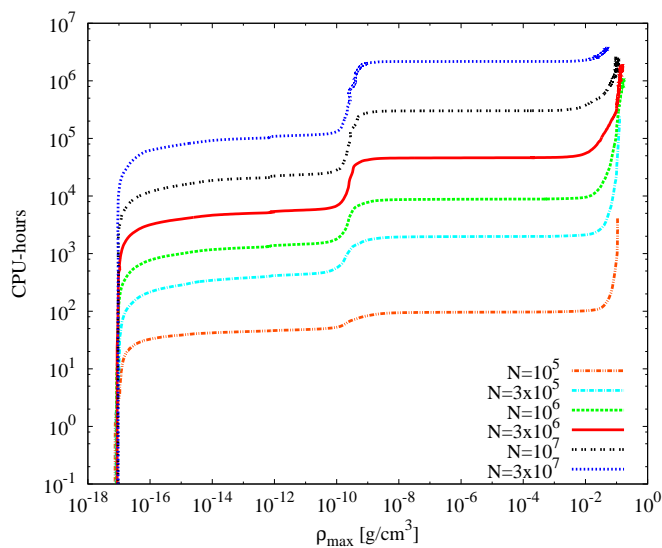}
\caption{Cumulative number of CPU hours per simulation as a function of maximum density, which is a proxy for time.   All simulations were performed on the DiRAC2.5 Data Intensive service at Leicester computer cluster using the hybrid openMP-MPI version of \textsc{sphNG}$^{\ref{foot1}}$.  The wall-clock equivalent for the \mtsv{} model is \sm2~yrs.}
\label{fig:cpuVrho}
\end{figure}

\figref{fig:cpuVrho} shows the number of CPU hours it takes to reach any given maximum density, which is a proxy for time.   Given our computational resources, the \mtsv{} model ended after approximately 2~yrs of wall-clock time using 256 CPUs; by the end of this simulation, we were modelling the stellar core in nearly real time.   Given that runtime increases super-linearly with resolution, even higher resolutions are currently prohibitively expensive to run.  All simulations were run using the hybrid openMP-MPI version of \textsc{sphNG}\footnote{\label{foot1}The \mf{} model was run only using openMP.} on the DiRAC2.5 Data Intensive service at Leicester computer cluster. 

\section{Results}
\label{sec:results}

As with our previous studies, we follow the gravitational collapse of the cloud core through the first hydrostatic core phase (\rhoxrange{-12}{-9}), through the second collapse phase (\rhoxrange{-8}{-4}) and into the stellar core phase, where we continue to evolve the models for 4 to 40~yrs, depending on the resolution (third column of \tabref{table:resolutions}).  We define the stellar core formation to occur at \dtsczero{}, which occurs when \rhoxeq{-4}.

\subsection{Evolution of the density and temperature}
\begin{figure} 
\centering
\includegraphics[width=\columnwidth]{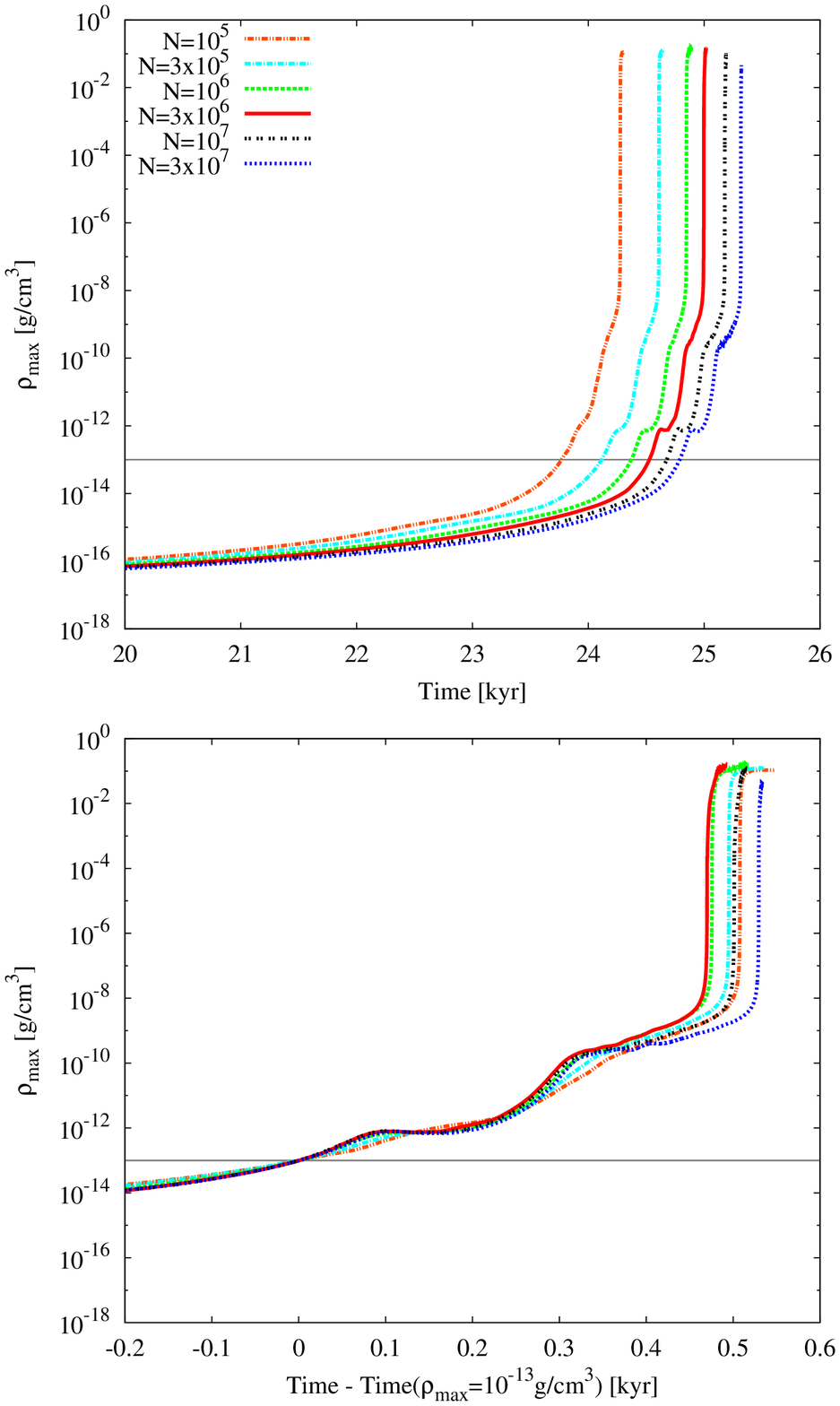}
\caption{Maximum density as a function of time for each model.  The top panel shows the absolute time, while the bottom panel shows the time normalised to the time when \rhoxeq{-13}, which is when the collapse exits the isothermal collapse phase; see the grey reference line.  There is a small `bounce' at \rhoapprox{-12} as the first core forms.  The models are reasonably converged when normalised to \rhoxeq{-13}.  This indicates that gravity during the isothermal collapse phase is the primary cause for the large range of absolute collapse times. }
\label{fig:RhoVtime}
\end{figure}

\begin{figure} 
\centering
\includegraphics[width=\columnwidth]{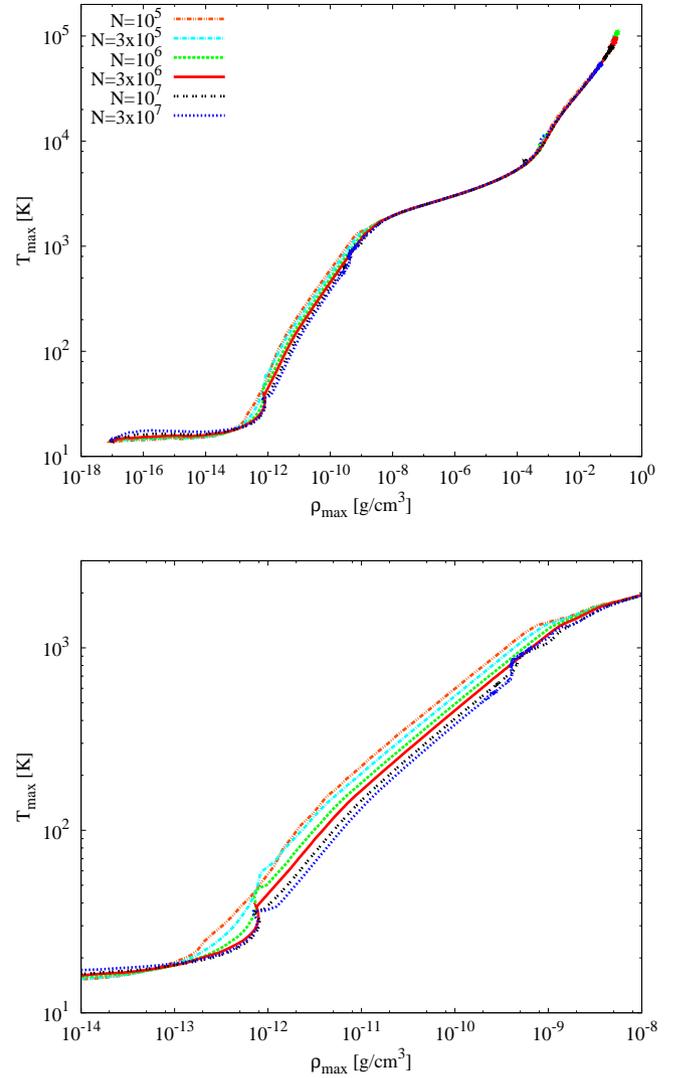}
\caption{Maximum temperature as a function of maximum density.  Top panel shows the entire evolution, while the bottom panel zooms in on the first core phase.  The bounce in maximum density near the formation of the first core (\rhoxapprox{-12}) is visible, where the maximum temperature continues to increase despite a slight overall cooling of the first core.  The heating at \rhoxtwoapprox{4}{-10} for \msvn{} is due to the formation of the magnetic wall.  The maximum temperature during the first core phase is slightly lower at any given \rhox{} for increasing resolution.  Outside of the first core phase, the maximum temperatures agree within 20 per cent.}
\label{fig:TvRho}
\end{figure}

In addition to physical mechanisms that delay the collapse of the cloud core \citep[as discussed by, e.g.,][]{BateTriccoPrice2014,Tsukamoto+2015oa,Tsukamoto+2015hall,\wpb2016,MachidaHiguchiOkuzumi2018,\wbp2018sd,\wbp2018ion,\wbp2018hd,\wbp2018ff,WursterBateBonnell2021}, numerical resolution also affects the overall collapse time \citepeg{WursterBate2019res}.  The top panel of \figref{fig:RhoVtime} shows the collapse times of our models, where the difference in reaching \rhoxeq{-4} is \sm1~kyr, with the lowest resolution collapsing the fastest.  By \rhoxapprox{-13}, the collapse exits the isothermal collapse phase and gas begins to trap radiation and heat up (see \figref{fig:TvRho}).  Thus, physical processes in addition to gravity become increasingly important.  When we normalise the collapse time to when each model reaches \rhoxeq{-13}, we find much better agreement, with the time between \rhoxrangeverb{-13}{-4} differing by only \sm60~yrs (bottom panel of \figref{fig:RhoVtime}).  Therefore, the discrepancy in collapse times primarily occurs during the isothermal collapse phase when gravity is the dominant physical process.  


As the first core forms at \rhoxapprox{-12}, there is a small `bounce' in the maximum density (\figref{fig:RhoVtime}).  During its formation, the core is essentially optically-thick and therefore behaves adiabatically during the bounce.  The core cools slightly, although the maximum temperature continues to slowly increase (\figref{fig:TvRho}).  The bounce is nearly non-existent in \mf{}, and is naturally better resolved for higher resolutions.

\begin{figure*} 
\centering
\includegraphics[width=\textwidth]{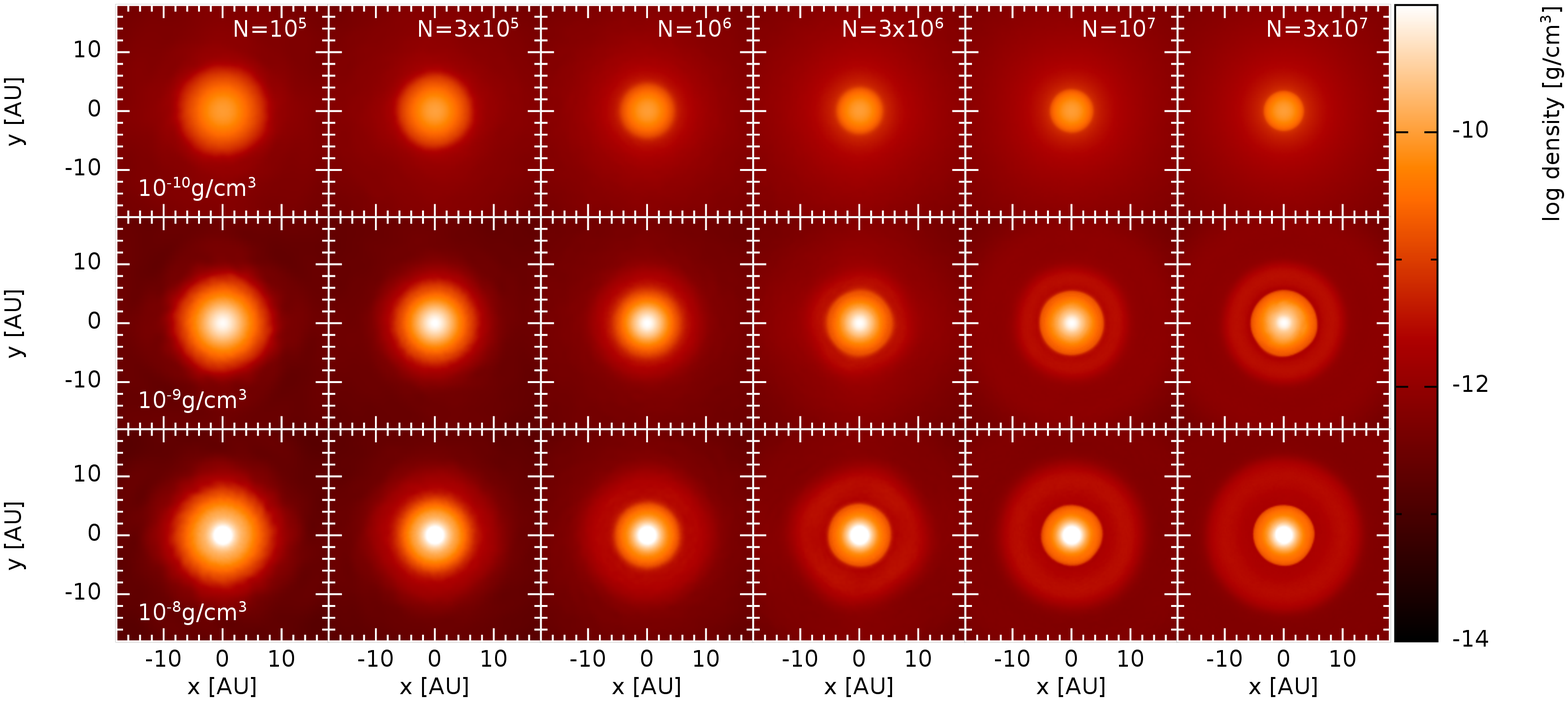}
\includegraphics[width=\textwidth]{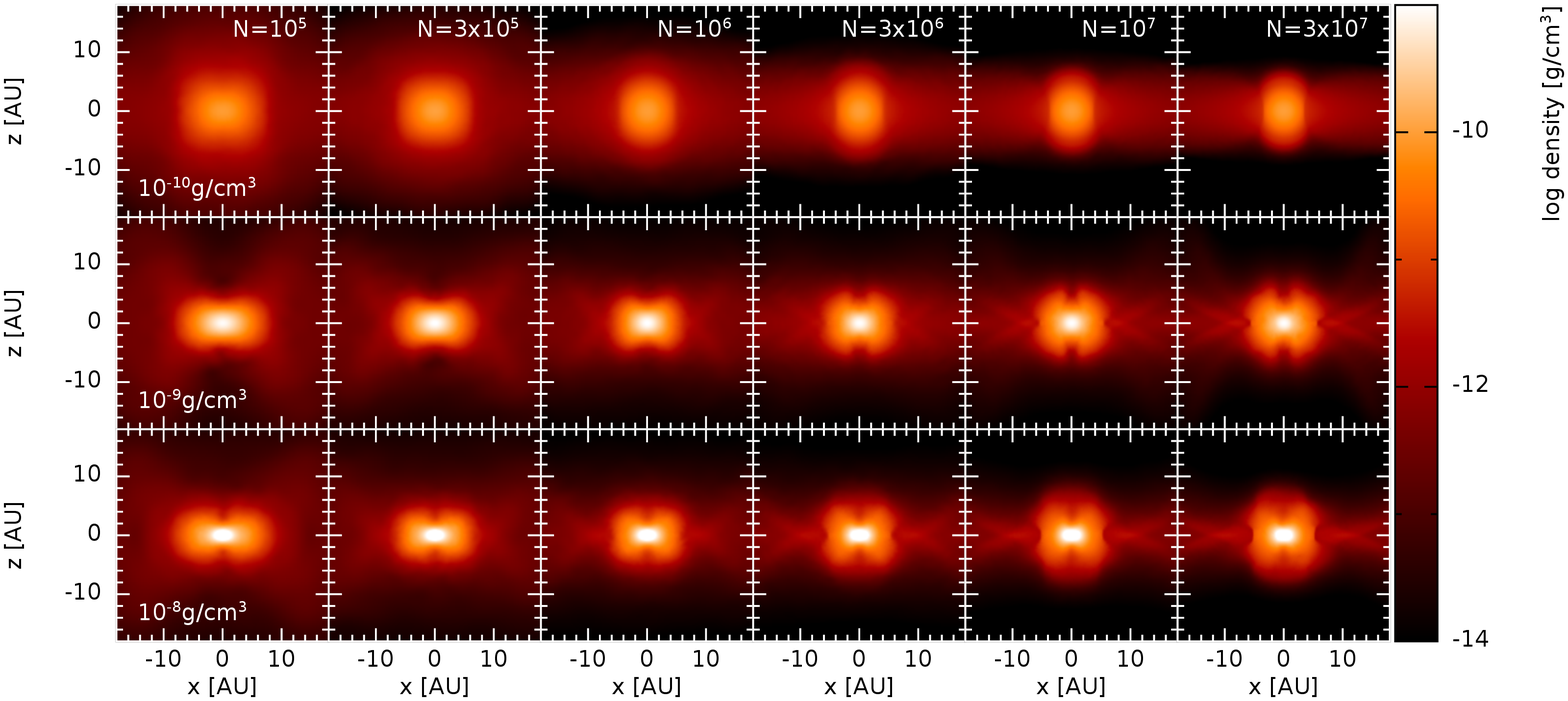}
\caption{Density slices through the first hydrostatic core perpendicular (top) and parallel (bottom) to the rotation axis for three maximum densities late in the first core phase.  Density profiles are broader at lower resolutions and the cores tend to be oblate; at higher resolutions, the density profiles are steeper and the first cores are prolate.}
\label{fig:fhc:rho}
\end{figure*}

\figref{fig:fhc:rho} shows the gas density at three times during the first core phase, both perpendicular and parallel to the rotation axis.  The high-density central regions of the core are similar at all resolutions (bottom row in each panel), however, the surrounding gas structure is resolution-dependent.  At low resolutions, radius of the first core is overestimated due to the comparatively large SPH smoothing length.  With increasing resolution, steeper density profiles are resolved, and the first core converges to a radius of $\approx 5$~au.  The shape of the core transitions from oblate to prolate as resolution is increased.  For \mtsxn{}, a distinct pseudo-disc forms in the mid-plane with density \rhogs{-12} around the prolate core, leading to a more structured core and central regions.

The lack of a consistent trend with resolution in the normalised collapse times (bottom panel of \figref{fig:RhoVtime}) is a result of the different gas structures in the first core and how well-resolved the bounce is.  When we consider only cores with similar structures (i.e. \mtsxn{}), then we again see a slight increase in collapse time with increasing resolution.  

\begin{figure*} 
\centering
\includegraphics[width=\textwidth]{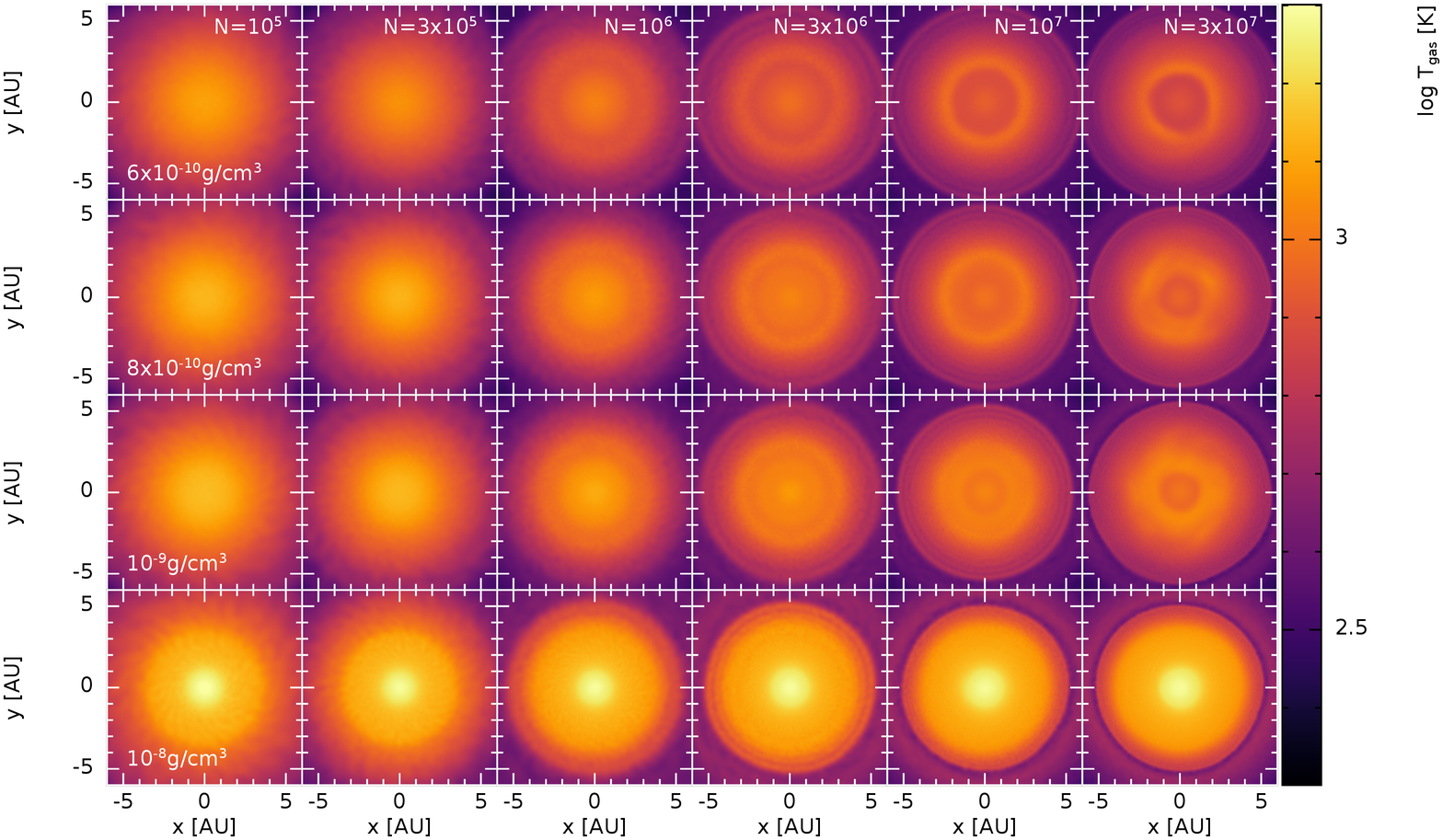}
\includegraphics[width=\textwidth]{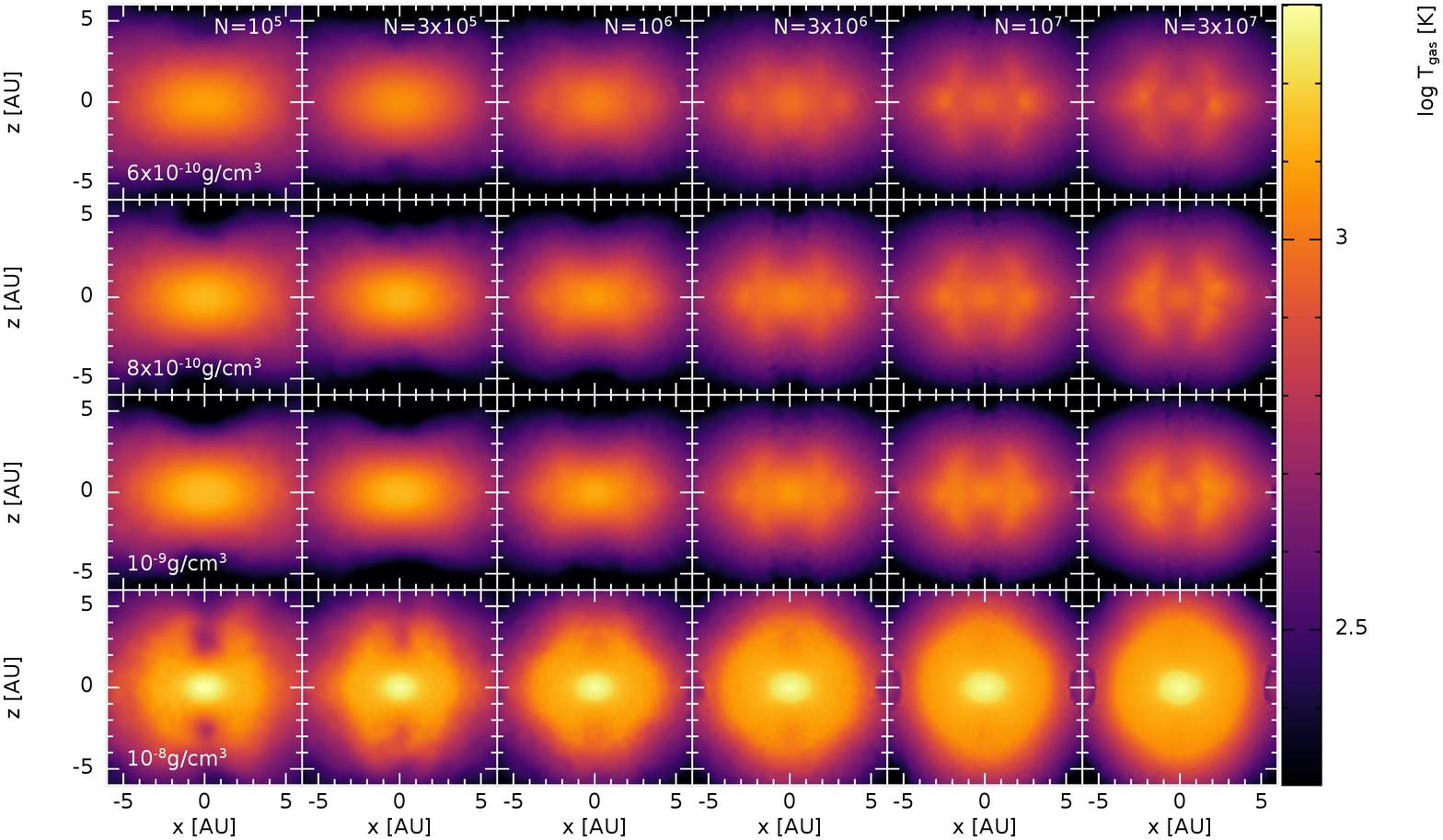}
\caption{Temperature slices through the first hydrostatic core perpendicular (top) and parallel (bottom) to the rotation axis for four maximum densities late in the first core phase; maximum densities and spatial scales are chosen to highlight our discussion of the magnetic wall in \secref{sec:MagWall}.  
At lower resolutions, the cores are hotter since the slightly larger cores are less efficient at radiating energy cool more slowly.  By \rhoxeq{-9}, the central region efficiently traps energy and heats up, nearly independent of the surrounding environment or resolution.}
\label{fig:fhc:T}
\end{figure*}

\figref{fig:fhc:T} shows the gas temperature at four times during the first core phase, both perpendicular and parallel to the rotation axis.  At low resolutions (\mtfx{}) or high maximum densities (\rhoxeq{-8}), the temperature profile in the equatorial plane (top panel) follows the density profile.   In the remaining panels, the temperature profile contains sub-structure and instead follows the magnetic field strength profile, which includes the magnetic wall (see \secref{sec:MagWall} below).  The vertical temperature profile (bottom panel) approximately follows the density profile for \rhoxle{-9}.  At higher maximum densities, the central core is more efficient at heating the surrounding gas to create the hot, spherical envelope.  

Prior to the formation of the first core, the maximum temperatures in the isothermal collapse phase (\rhoxls{-13}) agree within 20 per cent. The variation in maximum temperature with resolution during the first core phase (\figsref{fig:TvRho}{fig:fhc:T}) is again due to the fact that steeper gradients can be resolved with higher resolution.  First, as we have seen above, at low resolution the size of the first core is over-estimated, meaning that radiation finds it more difficult to escape.  Second, in the flux-limited diffusion approximation, the radiative flux is proportional to the gradient of the radiation energy density, which can be steeper with higher resolution, leading to more rapid cooling.  The bounce of the first core can also be seen in \figref{fig:TvRho} at \rhoxapprox{-12}, and the heating due to the formation of the magnetic wall can be seen at \rhoxtwoapprox{4}{-10} in the high resolution calculations.  After the first core phase, the maximum temperature is no longer in the magnetic wall and becomes dependent only on the central gas that is collapsing to ultimately form the stellar core; during this second collapse phase, the maximum temperatures agree within five per cent amongst the resolutions.  

\begin{figure*} 
\centering
\includegraphics[width=\textwidth]{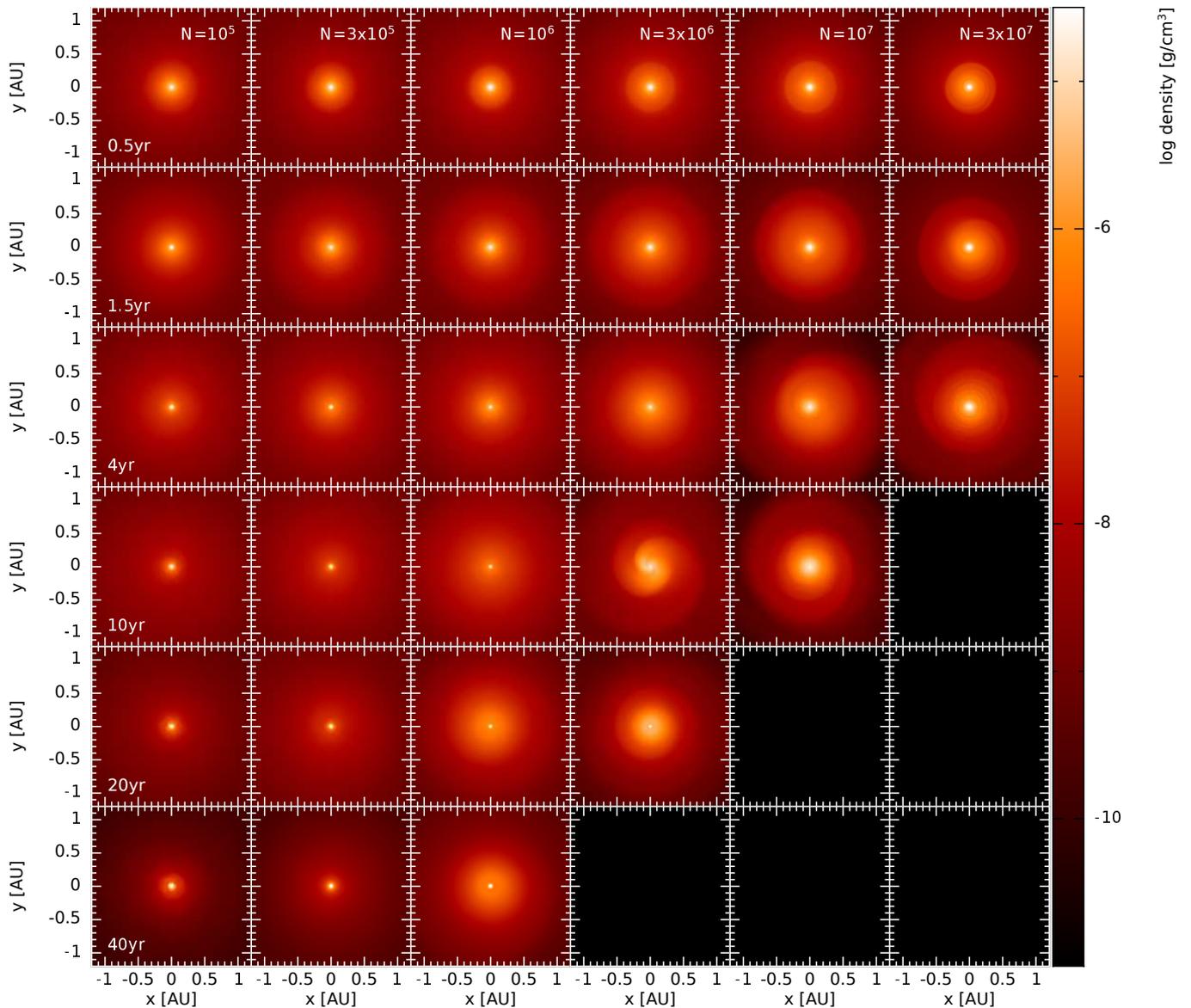}
\caption{Density slices through the stellar core perpendicular to the rotation axis as measured from the formation of the stellar core at \rhoxeq{-4}.  As the systems evolves, the higher resolution models (\msxn{}) form small discs that develop gravitational instabilities (e.g. the $m=2$ instability at 10~yr for the model with \mtsx{}) while the lower resolution simulations retain smooth density profiles that steepen with time.}
\label{fig:shc:rho}
\end{figure*}

\begin{figure*} 
\centering
\includegraphics[width=\textwidth]{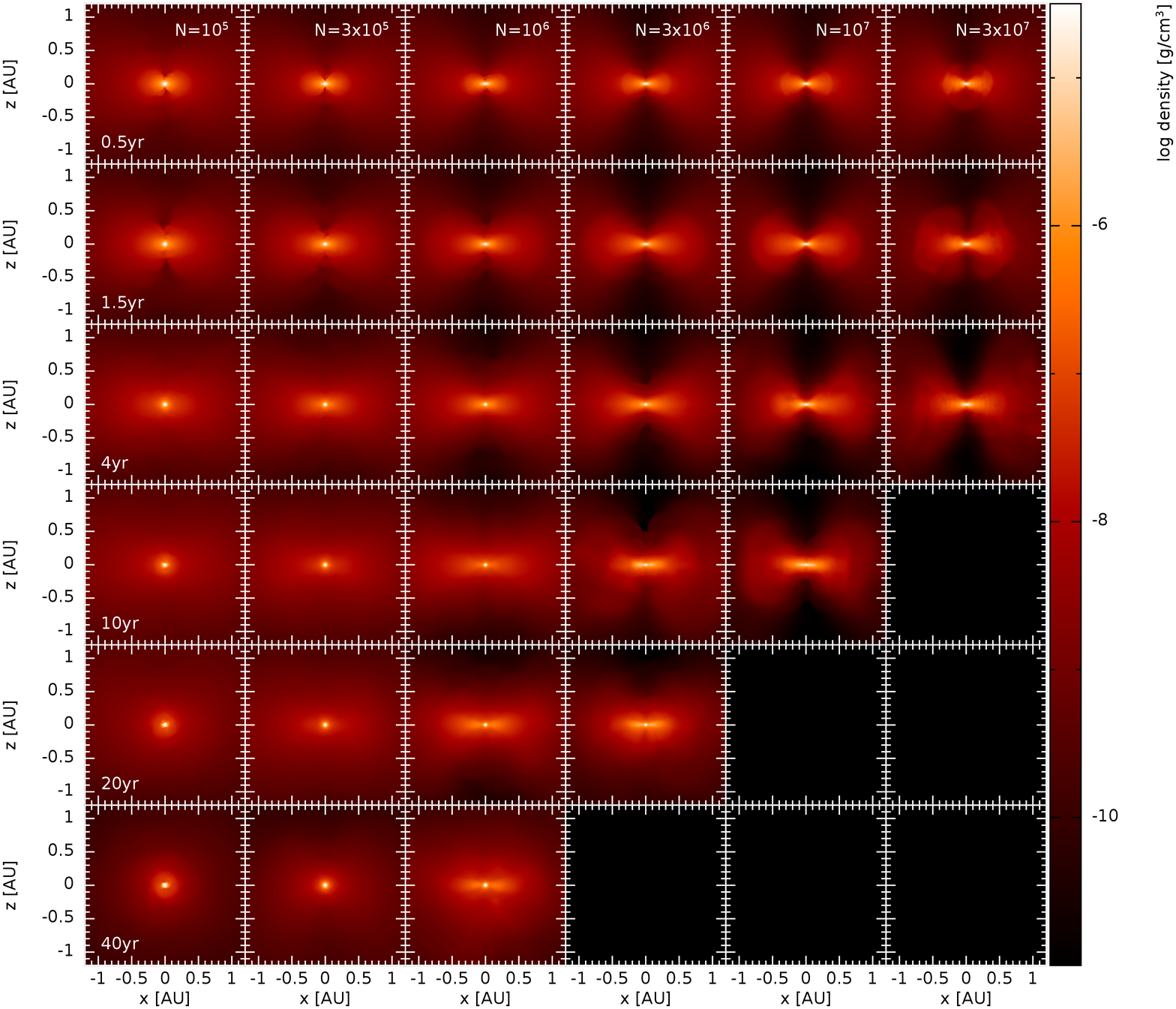}
\caption{Density slices through the stellar core as in \figref{fig:shc:rho}, but parallel to the rotation axis.  The flattened discs form and persist for resolutions of \msxn{}, while they dissipate for \mtfx{}.}
\label{fig:shc:rho:xz}
\end{figure*}

After the formation of the stellar core at \rhoxeq{-4}, it is more reasonable to compare the models normalised to the formation time of the stellar core, \dtsczero{}.   \figsref{fig:shc:rho}{fig:shc:rho:xz} shows the gas density around the stellar core for several times after its formation, which is resolution-dependent.  In all cases, the gas surrounding the stellar core has a disc-like distribution immediately after stellar core formation (top two rows of \figref{fig:shc:rho:xz}).  However, this is quickly lost with \mtfx{}, probably due to rapid angular momentum transport caused by SPH artificial viscosity, but also magnetic torques.
At higher resolutions, small, $r \lesssim 1$~au discs form and persist until the end of the simulations.  Furthermore, the discs in \mtsxn{} form weak spiral instabilities, with the most notable being a prominent but transient $m=2$ instability at 10~yr for \mtsx{}.  The stellar cores themselves are hot ($T \sim 10^4 - 10^5$~K), while the surrounding gas is much cooler ($T \sim 3000$~K). 

\subsection{Evolution of the magnetic field before stellar core formation}
\label{sec:Bfieldevol:FHC}

\subsubsection{Growth of the magnetic field strength}

\begin{figure} 
\centering
\includegraphics[width=\columnwidth]{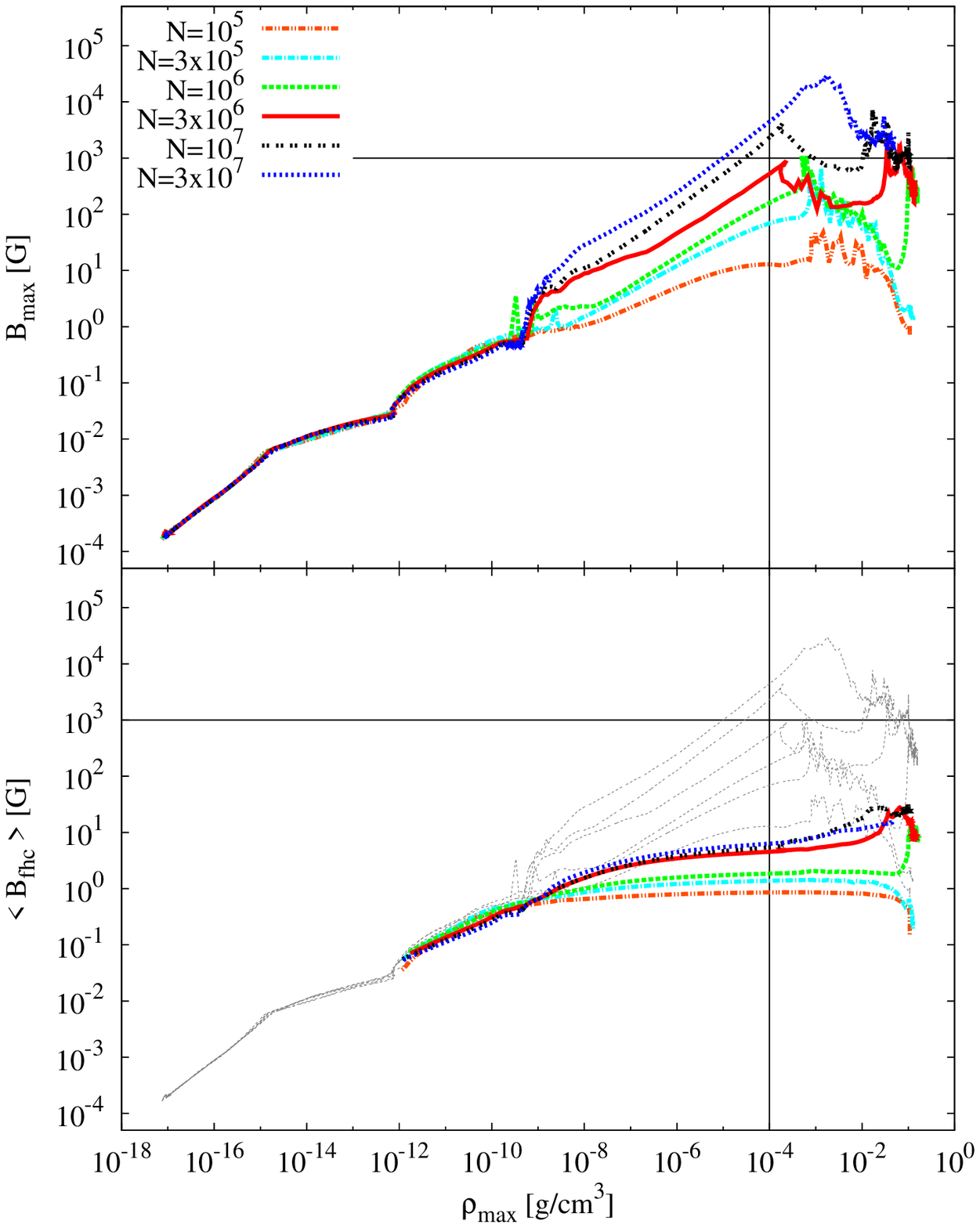}
\caption{Evolution of the maximum magnetic field strength against maximum density (top), and the evolution of the average magnetic field strength in the first hydrostatic core (bottom).  The vertical reference line represents the formation of the stellar core, the horizontal reference line represents the observed threshold of 1~kG, and the grey lines in the bottom panel are the maximum field strengths copied from the top panel for reference.  The average field strength is calculated as $\left<B_\text{fhc}\right> = 10^{\left[\sum_i \log(B_i)\right]/n}$, and includes only gas with \rhoge{-12} (i.e. the gas in the first or second hydrostatic cores).  The maximum magnetic field strength increases with increasing resolution starting late in the first core phase; the maximum field strengths in our higher resolution models surpass the 1~kG threshold used to determine the origin of magnetic fields in low-mass stars.  The average magnetic field strength typically remains below 10~G, and is approximately converged for \mtsxn{}. }
\label{fig:BVrho}
\end{figure}

\figref{fig:BVrho} shows the evolution of the maximum and average magnetic field strengths (i.e. \Bmax{} and \Bfhc{}, respectively).  The maximum magnetic field strength is approximately independent of resolution until late in the first core phase (\rhoxapprox{-9}); we have previously shown that it is in the early first core phase (\rhoxapprox{-12}) where the non-ideal processes cause \Bmax{} to diverge from ideal MHD simulations, with the growth rate slower in the models employing non-ideal MHD \citep{\wbp2018sd,\wbp2018ff,WursterBateBonnell2021}.  The sudden increase in \Bmax{} at the end of the first core phase (\rhoxapprox{-9}) is from the formation of the magnetic wall (see \secref{sec:MagWall} below), where the magnetic field is `piling up'; there is no corresponding increase in \Bfhc{}, indicating that the increase in \Bmax{} is localised to the wall and not spread throughout the first core.

By the end of the first core phase (\rhoxapprox{-8}), the maximum magnetic field strength differs by factor of \sm30 amongst the resolutions, but the average magnetic field strength is well converged for resolutions \mtsxn{}.  By the formation of the stellar core at \rhoxeq{-4}, the maximum field strength differs by \sm300, although this difference is exacerbated by the slower growth rate of \mf{} compared to the remaining models; this difference decreases to a factor of \sm65 when excluding \mf{} and to a factor of \sm10 when including only the three highest resolutions.  This increasing magnetic field strength for increasing resolution is due to both the gas being better resolved and less numerical dissipation from artificial resistivity.   Therefore, although our models are converging, we have not yet reached convergence in \Bmax{}.
  
At \rhoxeq{-4}, the maximum field strength of the models with \msvn{} reaches the observed \sm~kG field strength of young, low-mass stars.  However, the maximum field strength resides outside the stellar core itself (\secref{sec:MagWall}).  When we consider the average magnetic field strength of the gas comprising the first core (i.e. \rhoxge{-12}, which includes the magnetic wall), then all the average field strengths prior to the formation of the stellar core are below 10~G.  

\subsubsection{Magnetic field in the first core}
\begin{figure} 
\centering
\includegraphics[width=\columnwidth]{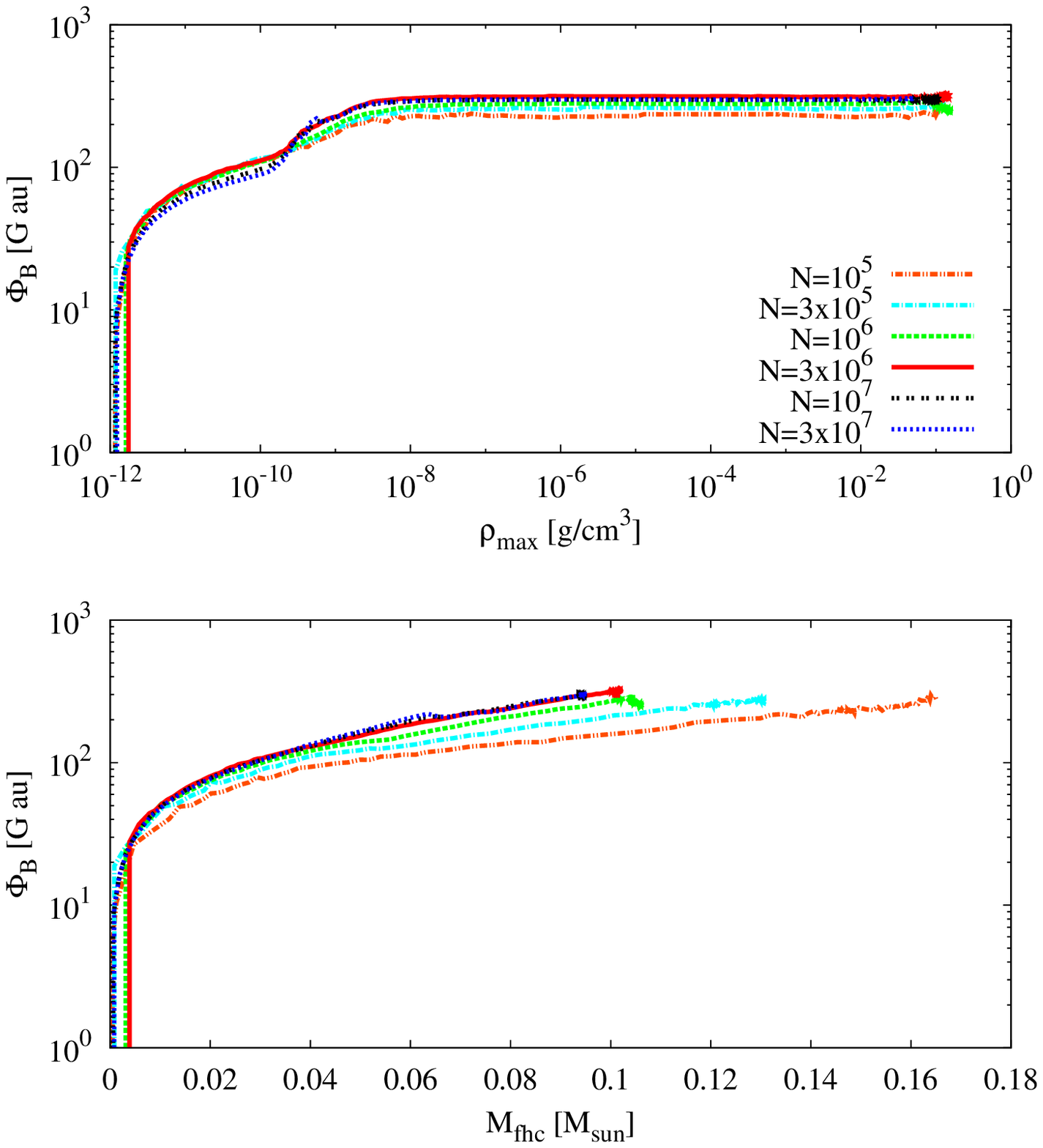}
\caption{Evolution of the magnetic flux passing through the mid-plane of the first hydrostatic core, $\Phi_\text{B} = \int_0^{r_\text{fhc}} B_\text{z} \text{d}A$.  Plotting against the mass of the first core (bottom) highlights the growth of the first core at \rhoxrange{-12}{-9}.  The increasing magnetic flux represents the magnetic field diffusing into the first core and the increasing magnetic field strength.}
\label{fig:Bflux:FHC}
\end{figure}

\figref{fig:Bflux:FHC} shows the magnetic flux threading the equatorial plane of the first hydrostatic core (i.e. the gas with \rhoxge{-12})\footnote{Note that this includes the pseudo-disc for \mtsxn{}.}.  As a function of \rhox{}, the magnetic flux is the same within 50 per cent for all models.  For \rhoxrange{-12}{-9}, the magnetic flux grows more rapidly than the radius, indicating that the magnetic field is being dragged in as the cloud gravitationally collapses and that the magnetic field strength is increasing.  This is confirmed by plotting the magnetic flux against mass of the first core (bottom panel of \figref{fig:Bflux:FHC}) during the first core phase (\rhoxrange{-12}{-9}); the magnetic flux increases as the mass of the first core increases, indicating that as mass enters the first core it drags the magnetic field with it.  

\subsubsection{The magnetic wall}
\label{sec:MagWall}
Unlike the density profile in the equatorial plane (top panel of \figref{fig:fhc:rho}), the magnetic field strength is not a smooth profile in the first hydrostatic core.  Instead, magnetic walls form as the magnetic field decelerates the charged particles, but not the neutral gas \citepeg{TassisMouschovias2005b,TassisMouschovias2007a,TassisMouschovias2007b}.  This prevents the magnetic flux from reaching the central regions and instead the magnetic field `piles up' in a torus of higher magnetic field strength such that the maximum magnetic field strength lies in this torus rather than at the centre of the core which is coincident with the maximum density.   This explains the rapid increase in \Bmax{} but not \Bfhc{} shown in \figref{fig:BVrho} since the increase is localised to the torus.

\begin{figure*} 
\centering
\includegraphics[width=\textwidth]{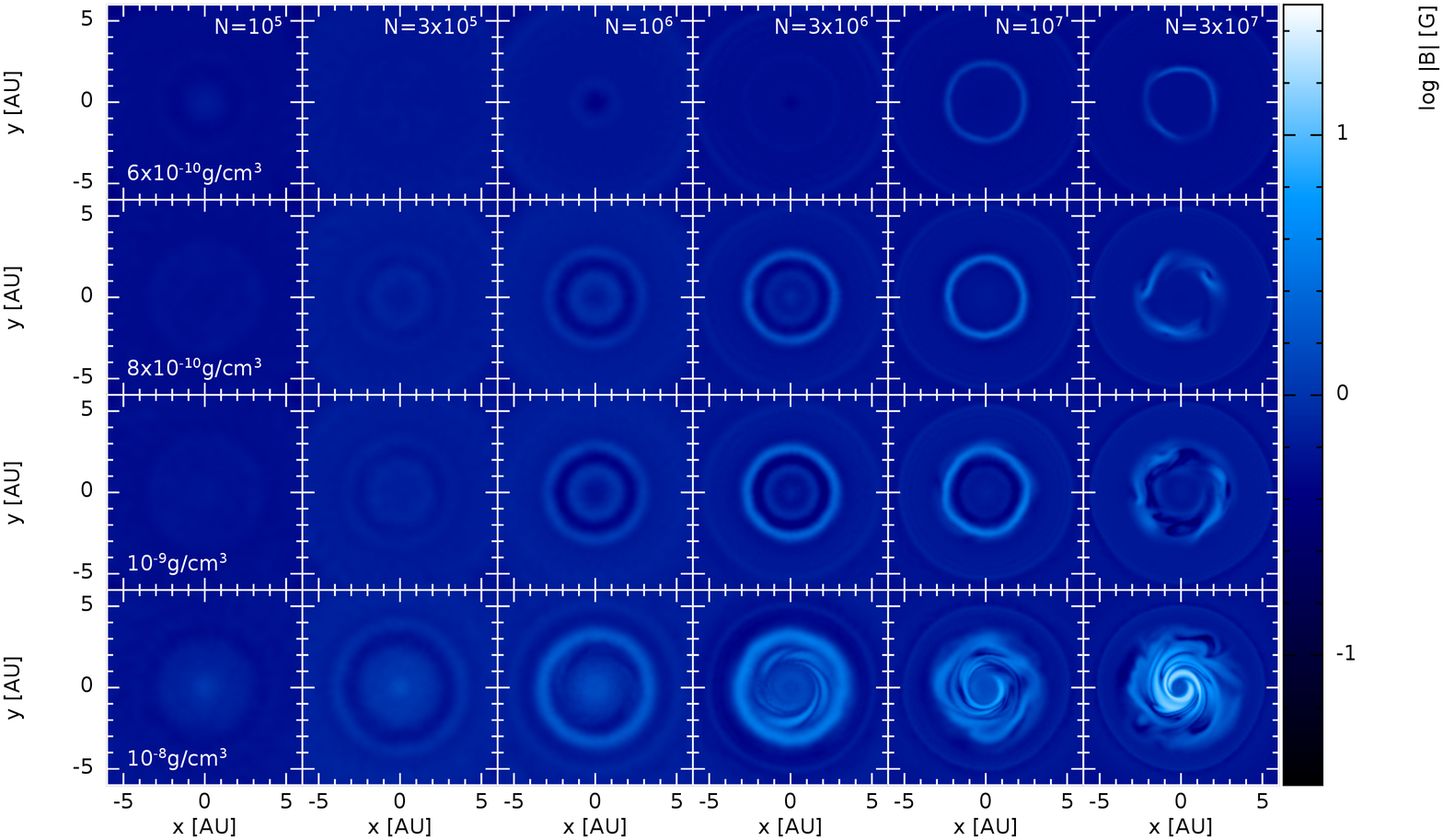}
\includegraphics[width=\textwidth]{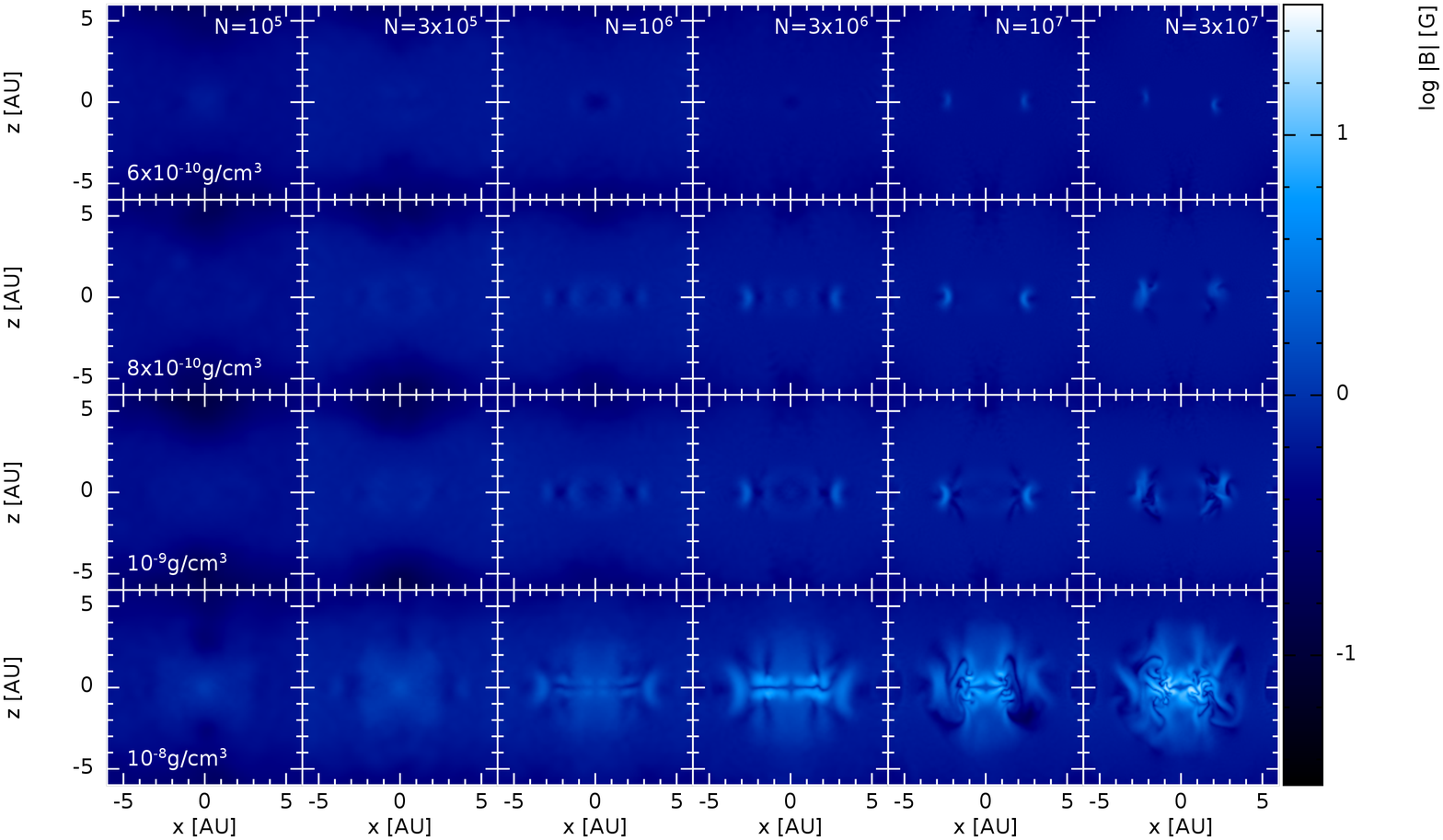}
\caption{Magnetic field strength slices through the first hydrostatic core perpendicular (top) and parallel (bottom) to the rotation axis for four maximum densities late in the first core phase; spatial range and maximum densities are chosen to highlight the formation and early evolution of the magnetic wall.  The magnetic wall forms earlier at higher resolutions, and is less axisymmetric.  The asymmetries are caused by the Hall effect, and the greater asymmetries at higher resolutions are a result of resolving shorter whistler waves.}
\label{fig:fhc:B}
\end{figure*}

\figref{fig:fhc:B} shows the magnetic field strength in slices through the core late in the first core phase, both perpendicular and parallel to the rotation axis.  The formation time of the magnetic wall is resolution-dependent, with the wall forming at slightly lower maximum densities at higher resolutions; this does not correspond to a large change in absolute or relative time (recall \figref{fig:RhoVtime}).  

The temperature profile (\figref{fig:fhc:T}) is similar to the magnetic field strength profile (\figref{fig:fhc:B}).  The magnetic walls are hotter than the surrounding gas, and the maximum temperature lies within this wall for \msvn{} between $\rho_\text{max} \approx 4\times 10^{-10}$ and $10^{-8}$~\gpercc{}.  Although there is a temperature increase in the walls at lower resolutions compared to the surrounding gas, the temperature in the walls is still lower than the central temperature for \mtsxx{}.  Both Ohmic resistivity and ambipolar diffusion contribute to heating the gas \citepeg{WursterPriceAyliffe2014}, thus as the ionised gas is slowed down to form the wall, it also heats up, accounting for the higher temperature in the magnetic walls.

Ambipolar diffusion tends to be the process that produces the strongest magnetic walls; assuming the initial cloud is axisymmetric and that the Hall effect is excluded, then the magnetic wall will also be axisymmetric \citep{WursterBateBonnell2021}.  The magnetic field strengths are similar at all resolutions until the formation of the magnetic wall (see \figref{fig:BVrho}), however, as discussed above, the density profile is smoothed out at lower resolutions.  Therefore, the density at $r \approx 3$~au is slightly higher at lower resolutions, meaning that the effect of ambipolar diffusion is weaker.  This results in the later formation time of the magnetic wall at lower resolutions.  Once the wall has formed, the higher resolution models have stronger magnetic field strengths in the wall since the magnetic field is better resolved and there is less artificial dissipation.  This results in a stronger effect of ambipolar diffusion, amplifying the wall.    

The spiral structure (bottom row of the top panel in \figref{fig:fhc:B} for \mtsxn{}) in the magnetic field strength is caused by the Hall effect \citep{WursterBateBonnell2021}.  Unlike Ohmic resistivity and ambipolar diffusion, the Hall effect is a dispersive term that splits the \alfven{} wave into left- and right-circularly polarised waves.  The right (whistler) wave propagates faster than the \alfven{} wave, and its speed increases for decreasing wavelength \citepeg{SanoStone2002a,PandeyWardle2008,WursterPriceBate2016,MarchandCommerconChabrier2018,Marchand+2019}.  Therefore, as numerical resolution is increased, additional whistler wavelengths are resolved.  This yields dispersion on smaller scales and results in the spiral structure in the magnetic field shown in \figref{fig:fhc:B}.   These additional whistler wavelengths prevent numerical convergence, at least at the resolutions presented here.

Between $\rho_\text{max} = 10^{-9}$ and $10^{-8}$ g~cm$^{-3}$, the gas in the centre of the first core becomes oblate (bottom panel of \figref{fig:fhc:rho}), and the magnetic field is amplified in this region (bottom panel of \figref{fig:fhc:B}).  By \rhoxeq{-8} for \msvn{}, the magnetic wall has been disrupted due to the short wavelength whistler waves, creating the highly structured magnetic field within the first core.  Despite this disruption, the maximum magnetic field strength continues to reside outside of the centre of the core. 

Therefore, ambipolar diffusion is primarily responsible for the formation of the magnetic wall, but the Hall effect is responsible for its asymmetric structure, with the Hall effect's importance increasing with increasing resolution as whistler waves with shorter wavelengths are resolved.

\subsubsection{The rapid second collapse}
\label{sec:Bfield2ndcollapse}

\begin{figure*} 
\centering
\includegraphics[width=\textwidth]{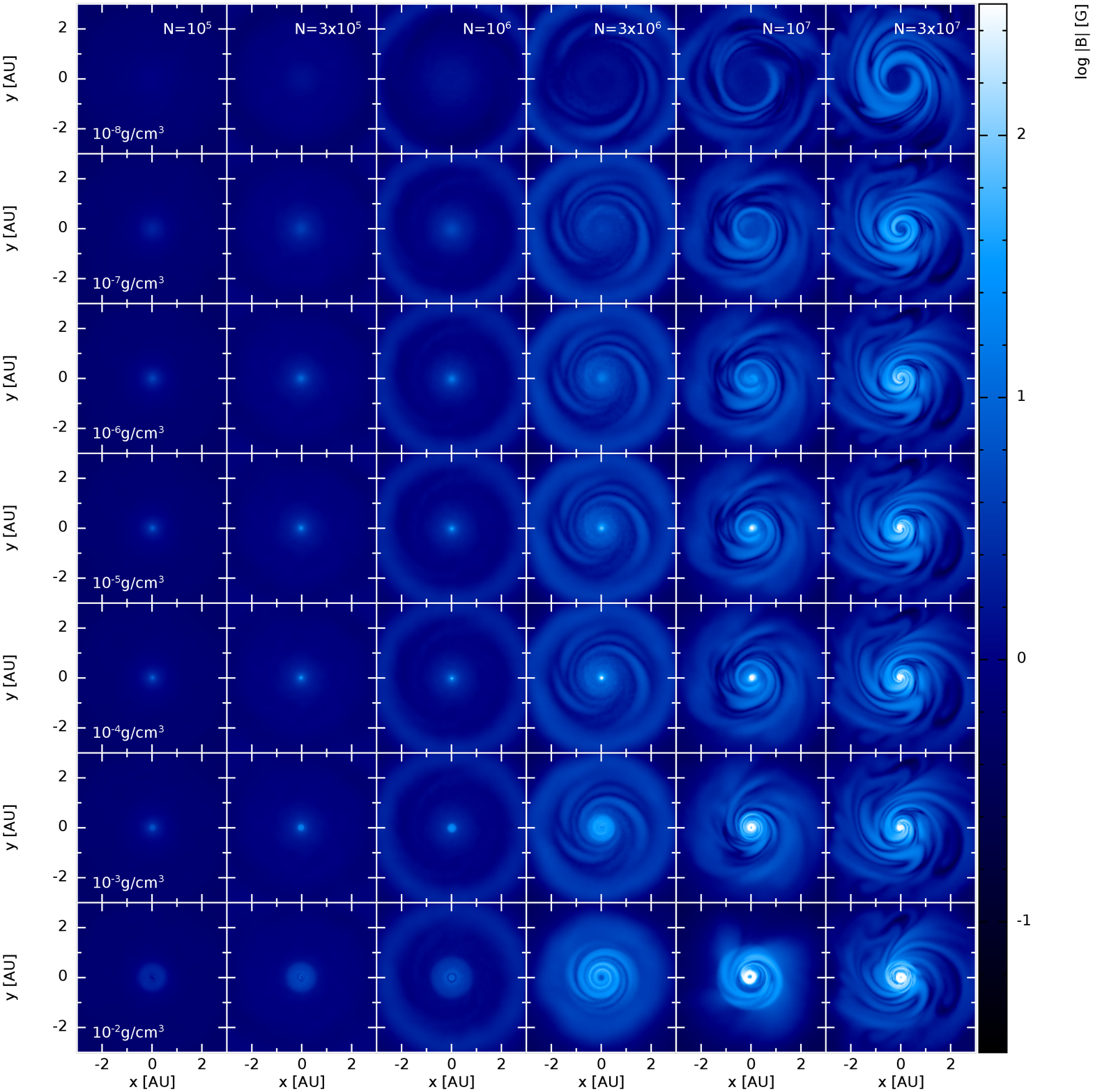}
\caption{Magnetic field strength slices through the inner parts of the first hydrostatic core that are perpendicular to the rotation axis during and after the rapid second collapse phase.  The collapse is so quick that the field does not evolve significantly outside $\approx 1$~au, but there is rapid growth on scales $\ll 1$~au.}
\label{fig:rapidcollapse:B}
\end{figure*}

\figref{fig:rapidcollapse:B} shows the magnetic field strength in slices through the inner parts of the first hydrostatic core during the rapid second collapse phase.  The collapse is so quick that the field does not evolve significantly outside $\approx 1$~au, but there is rapid growth on scales $\ll 1$~au.  In the higher resolution models, much of the magnetic flux that ends up in the vicinity of the stellar core comes from the highly-structured magnetic field that was produced by the Hall effect during the disruption of the magnetic wall.  This indicates that to obtain a complete understanding of how fossil magnetic fields are implanted into stellar cores may require accurate modelling of the Hall effect.

\subsection{Evolution of the magnetic field after stellar core formation}
\label{sec:Bfieldsc}

To determine the origin of magnetic fields in low-mass stars, we investigate the magnetic field strength at and immediately after the formation of the stellar core.   In \citetalias{\wbp2018ff} at a resolution of \mtsx{}, we concluded that the magnetic fields in low-mass stars were generated by a dynamo later in life since the stellar core magnetic field strengths at birth were $< 1$~kG.  

\begin{figure} 
\centering
\includegraphics[width=\columnwidth]{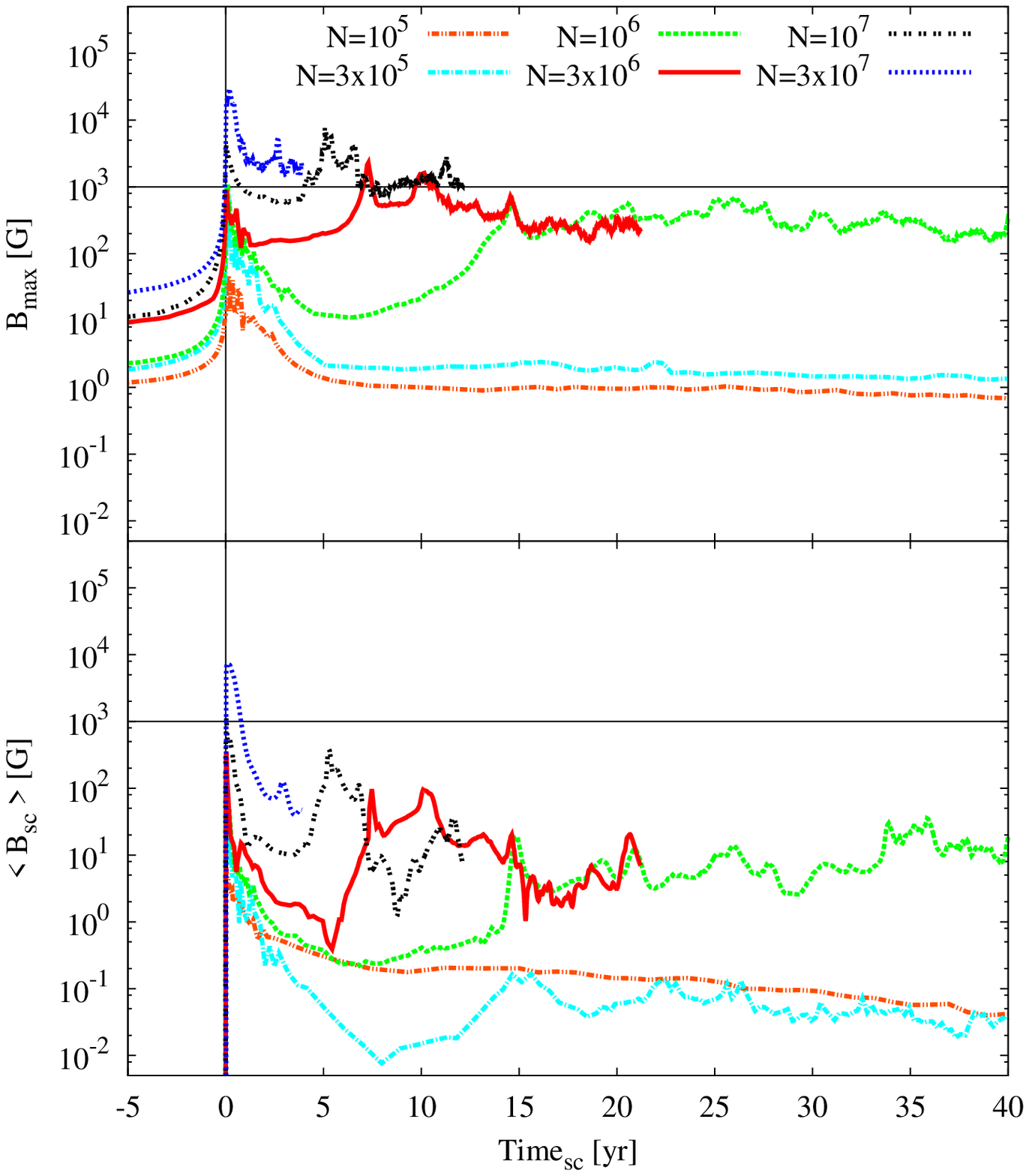}
\caption{Evolution of the maximum (top) and stellar core (bottom) magnetic field strengths, as measured from the formation of the stellar core.  The average field strength is calculated as $\left<B_\text{sc}\right> = \left[\sum_i \log(B_i)\right]/n$, and includes only gas with \rhoge{-4}.   Vertical and horizontal solid lines are included for reference.  For increasing resolution, both field maximum and stellar field strengths increase, where the maximum magnetic field strength resides outside of the stellar core.  Fluctuations in the stellar core strength result from magnetic flux being advected between the core itself and the surrounding gas.  For the first six months after stellar core formation in \mtsv{}, the stellar core strength surpasses the observed 1~kG threshold.}
\label{fig:B:SHC}
\end{figure}

\figref{fig:B:SHC} shows the maximum magnetic field strength, \Bmax{}, and average field strength of the stellar core, \Bsc{}, after its formation.  The maximum field strength in \mtsv{} surpasses the kG threshold at stellar core formation and remains above until the end of the simulation 4~yr later.  For models with \mtsxx{}, the maximum magnetic field strength is $B_\text{max} \lesssim 1$~kG.  Therefore, there is a resolution-dependence on the maximum magnetic field strength, and whether it is above or below the kG threshold.

However, as mentioned above in \secref{sec:MagWall}, the maximum magnetic field strength lies outside the stellar core itself.  The bottom panel of \figref{fig:B:SHC} shows the average magnetic field strength in the stellar core.  The average magnetic field strength in the stellar core in the first few years after its formation is similar for models with $N \leq 3\times 10^6$, and lies well below 1~kG.  However, it increases significantly for the two highest resolution models, and for the highest resolution simulation it exceeds 1~kG for approximately 6 months after stellar core formation.  

\begin{figure*} 
\centering
\includegraphics[width=\textwidth]{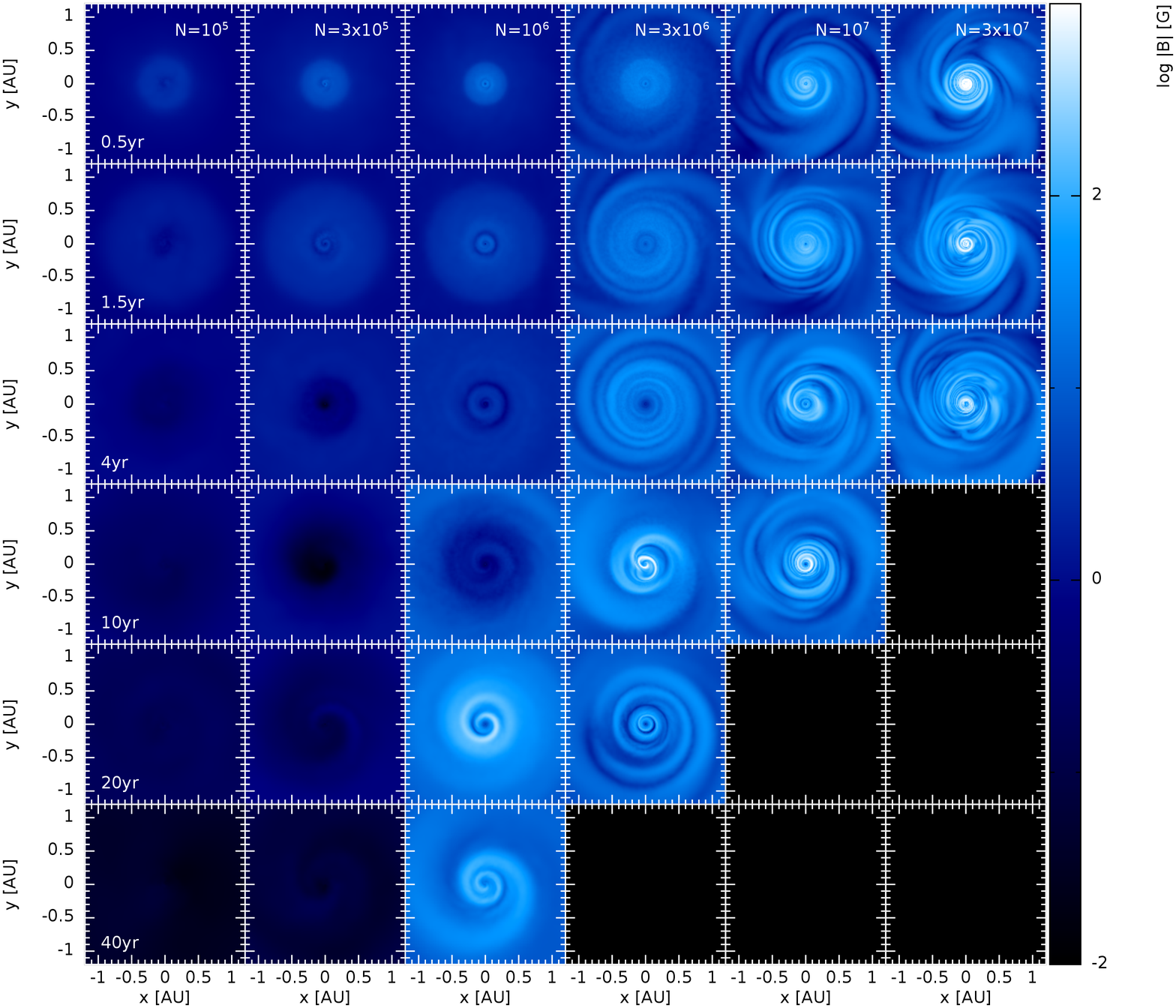}
\caption{Magnetic field slices through the stellar core perpendicular to the rotation axis. For increasing resolution, the magnetic field strength increases and the spiral structures become finer and more intricate.  }
\label{fig:shc:Bxy}
\end{figure*}

After stellar core formation, there is an immediate decline in magnetic field strength, followed by an evolution where the field strengths fluctuate up to \sm2~dex, with larger fluctuations in \Bsc{} than \Bmax{}.  Although the field strength decreases for \mtfx{}, the fluctuations and short evolution time make extracting a trend from the remaining models challenging.  Nonetheless, \figref{fig:B:SHC} shows that the magnetic field in the core is continuing to dynamically evolve.  \figsref{fig:shc:Bxy}{fig:shc:Bxy:small} show the magnetic field strength in a slice through the stellar core perpendicular to the rotation axis; the two figures show the field strengths on different spatial scales, and the defined boundary of the stellar core at \rhoxeq{-4} is shown in the latter figure.  
\begin{figure*} 
\centering
\includegraphics[width=\textwidth]{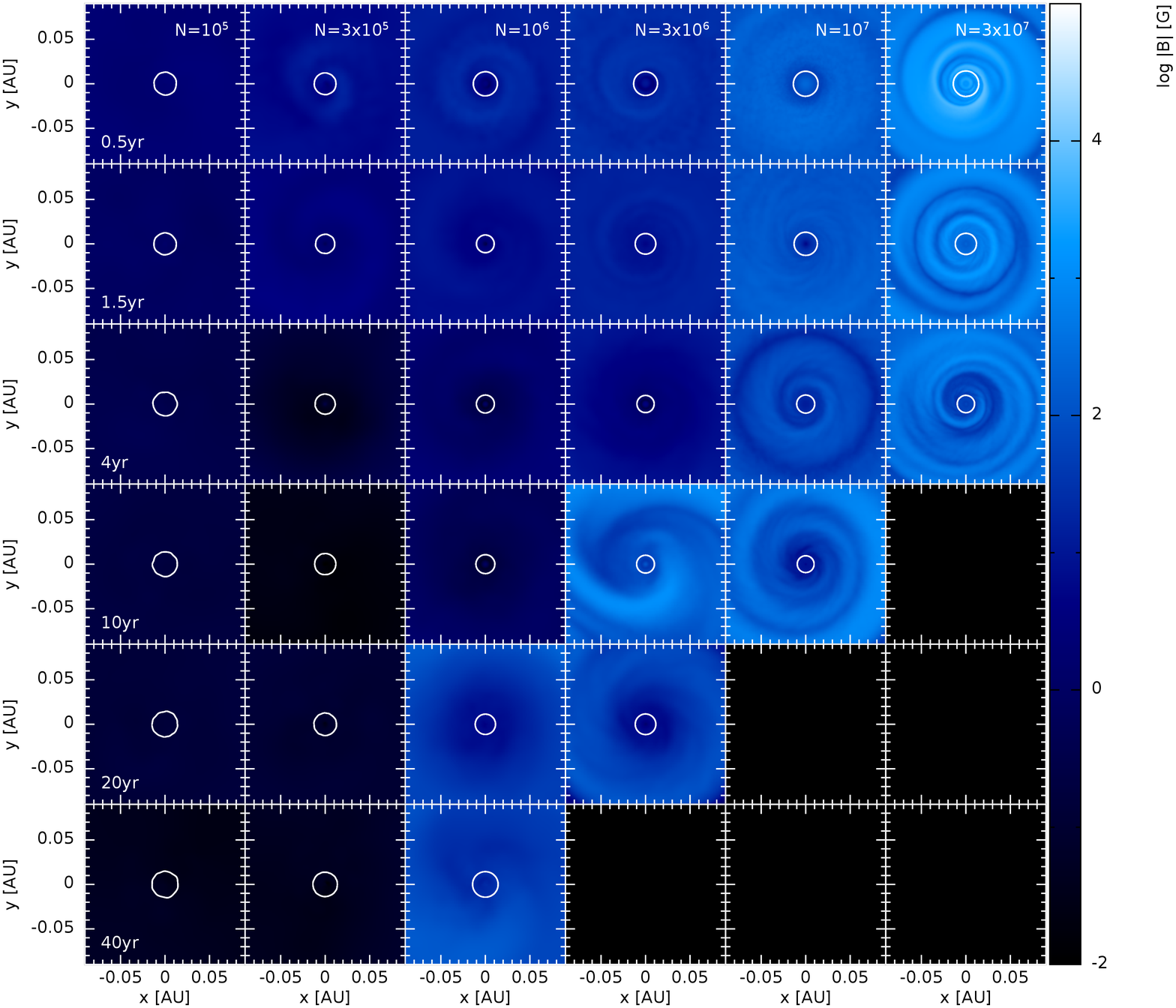}
\caption{Magnetic field slices through the stellar core as in \figref{fig:shc:Bxy}, but on a smaller spatial scale.  The contour represents the boundary of the stellar core at \rhoxeq{-4}.  The magnetic field is continually advected in and out of the core due to the continual gas distribution between the core and its surrounds, leading to fluctuations in the magnetic field strength of the stellar core shown in \figref{fig:B:SHC}.  }
\label{fig:shc:Bxy:small}
\end{figure*} 

Naturally, for higher resolution, numerical dissipation\footnote{At the temperatures in the stellar core, physical dissipation is negligible.} is lower and the magnetic structures are better resolved, as can be seen in these figures.  This leads to higher magnetic field strengths, as previously discussed.  At low resolutions (\mtfx{}), the entire region in and surrounding the stellar core is permeated with a weak, unstructured magnetic field.  For \msx{}, the field strength grows with time, with an increase of \sm1~dex at \dtscapprox{14}, which corresponds to the formation of the spiral structure clearly seen at \dtsc{20} in \figref{fig:shc:Bxy}; note that this spiral structure does not exist in the density profile, confirming that it is produced by \alfven{} waves.

Similar to the magnetic wall (\secref{sec:MagWall}), the magnetic field structures are well-defined for \mtsxn{}, with spiral structures seen in the strength of the magnetic field surrounding the stellar core (\figsref{fig:shc:Bxy}{fig:shc:Bxy:small}).  These structures are naturally tighter and more finely structured for increasing resolution.  This complex magnetic field is continually being advected between the core itself and the surrounding gas since, at this stage, there is no rigid boundary at the edge of the core.  Moreover, the evolution leads to transitions between the toroidal and poloidal components of the magnetic field within the core itself.  This evolution of the magnetic field in and near the core leads to the fluctuations shown in the bottom panel of \figref{fig:B:SHC}.  

Therefore, the magnetic field in and around the core is dynamically evolving, leading to fluctuations in both the maximum and stellar core magnetic field strengths.  Furthermore, these fluctuations are resolution-dependent, and our results are not numerically converged.  This clearly illustrates the difficulty of determining the strength of the fossil field that may be implanted in a stellar core. However, in our highest resolution model, the average magnetic field strength of the stellar core does exceed the 1~kG threshold for \dtscapproxmo{6}, suggesting that with even higher resolution the observed magnetic fields of young low-mass stars may be able to be provided by fossil fields.  The temperatures within the stellar core are high enough that artificial resistivity in the stellar core is likely responsible for the decay of the magnetic field strength after stellar core formation.

\subsection{First hydrostatic core outflows}

\begin{figure*} 
\centering
\includegraphics[width=\textwidth]{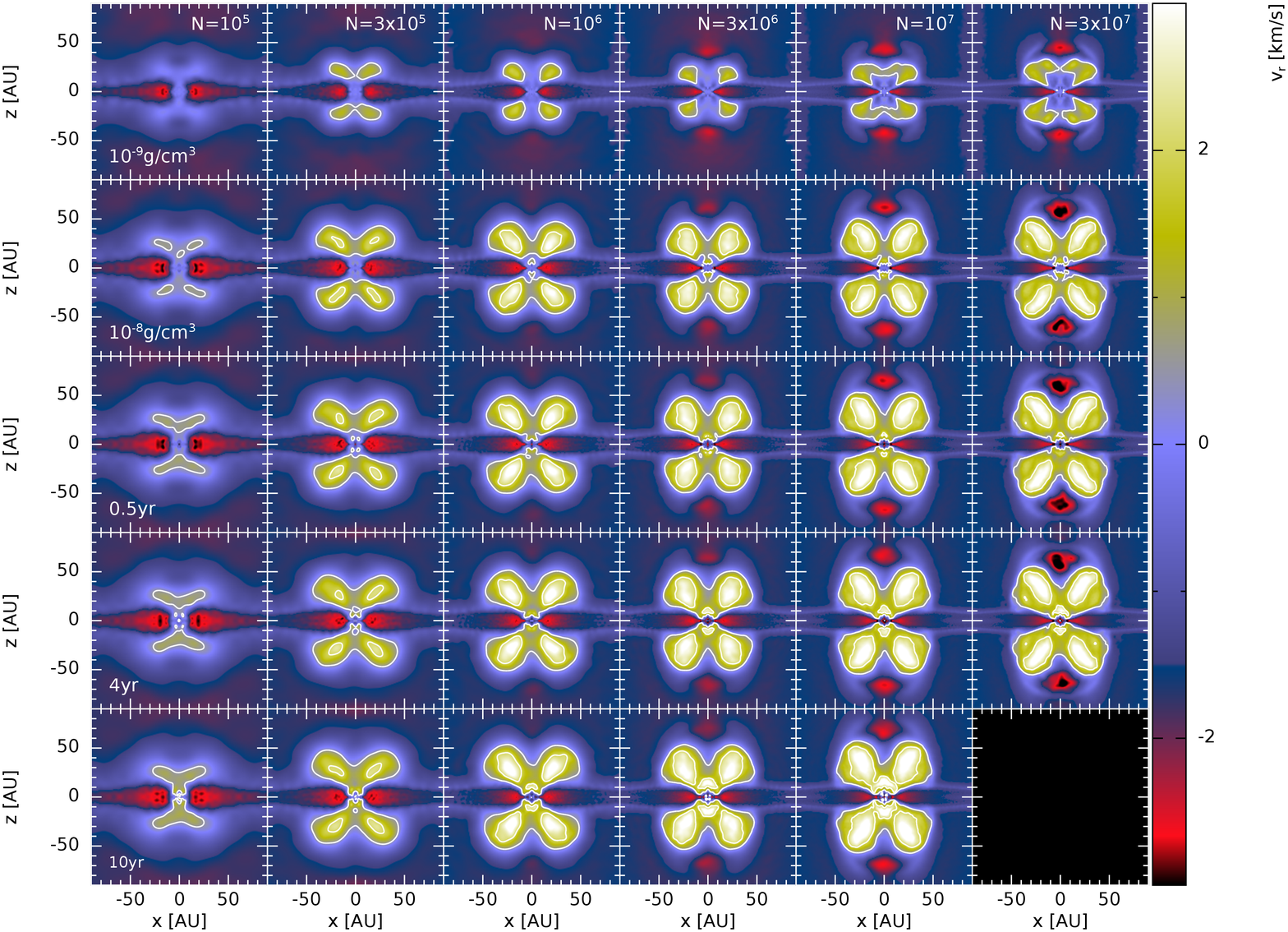}
\caption{Radial velocity of the first core outflow in a slice through the core perpendicular to the rotation axis.  The top two rows are plotted at constant maximum density late in the first core phase, while the bottom three rows are plotted after the formation of the stellar core, as measured from its formation.  Contours are at $v_\text{r} = 0.5$ and $2$~\kms{}, which are the boundary velocities for the slow and fast outflows shown in \figref{fig:outflow}.  The fast component of the outflows initially exists in the lobes, but at late times, an additional fast component forms above and below the core.}
\label{fig:vr}
\end{figure*}

\begin{figure} 
\centering
\includegraphics[width=\columnwidth]{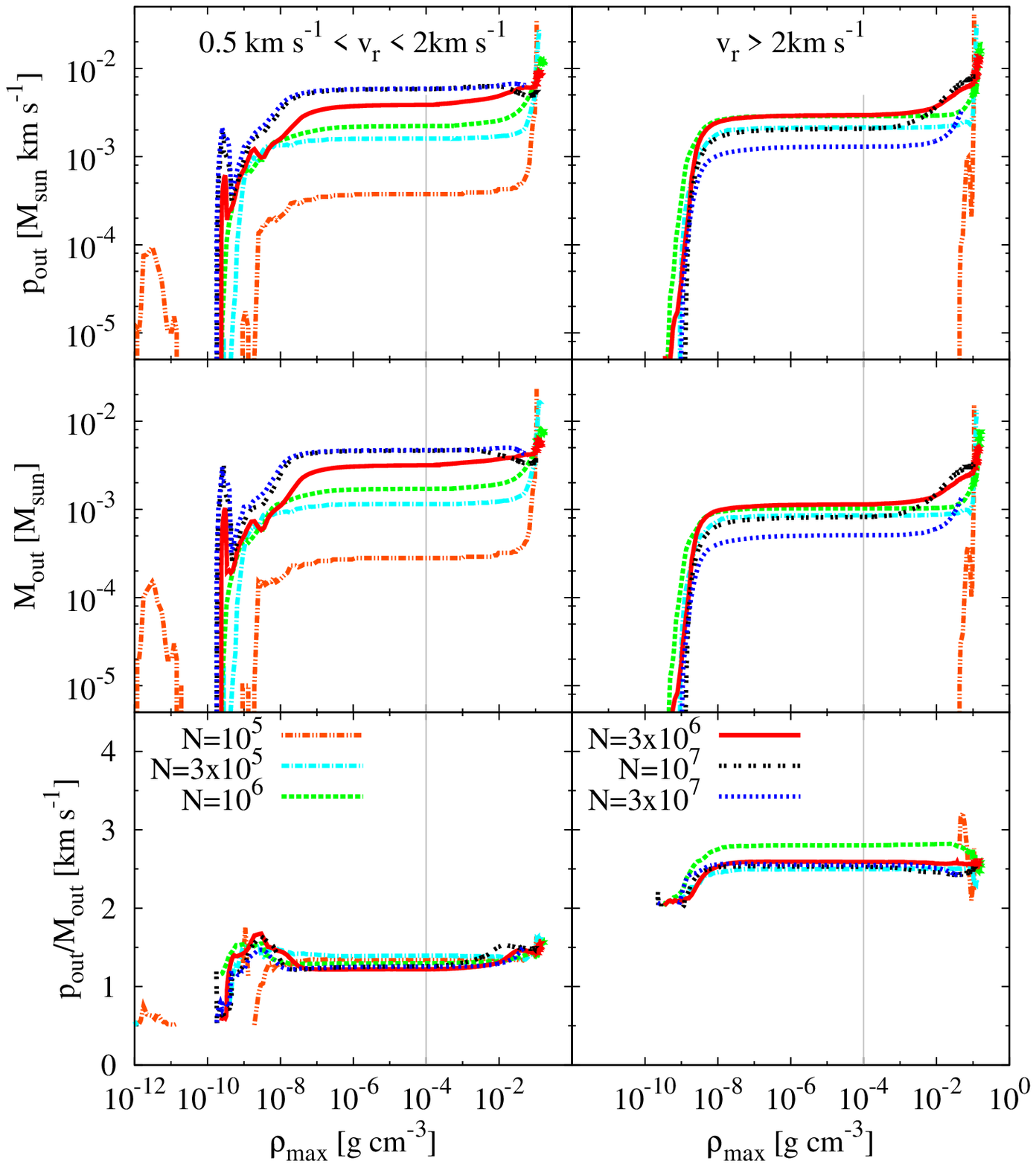}
\caption{The total momentum (top row), total mass (middle row) and average velocity (bottom row) in the slow ($0.5 < v_\text{r}/$(\kms)$ \ < 2$; left-hand column) and fast ($v_\text{r} > 2$~\kms; right-hand column) outflows. The vertical grey line represents the defined formation density of the stellar core.  Gas is in the outflow if it is at least 30$^\circ$ above/below the mid-plane, its radial velocity vector is at least 30$^\circ$ above/below the mid-plane, and satisfies \rhole{-8} and $|v_\text{r}|/|v| > 0.5$.  The slow outflow has converged for \msvn{}, while the fast outflow has not yet converged (excluding velocity), with mass and momentum slightly decreasing for increasing resolution.} 
\label{fig:outflow}
\end{figure}

First core outflows are magnetically launched from the pseudo-disc during the first core phase.  They typically expand at a few \kms{} and contain $M < 0.01$~\Msun{} by the formation of the stellar core \citepeg{\wpb2016,\wbp2018hd,WursterBateBonnell2021};  models that include all three non-ideal processes and are initialised with the rotation and magnetic field vectors aligned as modelled here yield the fastest outflows compared to other orientations or combination of non-ideal processes \citep{WursterBateBonnell2021}.  \figref{fig:vr} shows the radial velocity of the first core outflows, and \figref{fig:outflow} shows the evolution of the momentum, mass, and average velocity of the outflows.  In the latter, the gas is defined to be in the outflow if it is at least 30$^\circ$ above/below the mid-plane, its radial velocity vector is at least 30$^\circ$ above/below the mid-plane, and satisfies \rhole{-8} and $|v_\text{r}|/|v| > 0.5$; we divide the outflow into fast ($v_\text{r} > 2$~\kms{}) and slow ($0.5$~\kms{}$ < v_\text{r} < 2$~\kms{}) components.

The first core outflow is not properly resolved with \mf{} (\figref{fig:vr}); the total mass of the outflow measure during the second collapse phase is \sm$3\times10^{-4}$~\Msun{}  (\figref{fig:outflow}) which corresponds to only 30 SPH particles.  This was the resolution of our star cluster simulations \citep{\wbp2019}, confirming that the reason they were not observed in that study was due to the resolution limit.  

A resolved outflow forms at resolutions \mtfn{}, while the structure qualitatively converges for \msvn{} (\figref{fig:vr}).  There is both a fast and slow component to the outflow, with the average velocity of each outflow well-converged at \sm1.5 and 2.5~\kms{} for the slow and fast outflows, respectively; for \msxn, part of each outflow reaches speeds of $v_\text{r} \gtrsim 3$~\kms{} at \rhoxeq{-8}.

With increasing resolution, there is increasing mass and momentum in the slow outflows, until the values converge for \msvn{} (left-hand column of \figref{fig:outflow}); the mass in \mtsx{} is only \sm1.5 times lower than the converged value, suggesting it also captures the outflow quite well.  The fast outflows are embedded in the lobes of the slow outflows, and there is decreasing mass and momentum for increasing resolution (when the fast outflows are resolved); during the second collapse, the mass in the fast outflow differs by only a factor of \sm2 amongst the resolutions.  Given that the fast component is part of the first core outflows rather than a separate outflow, there is some blurring between the fast and slow components, leading to this difference amongst the resolutions.  We conclude that first core outflows are well modelled for resolutions of \mtsxn{}. 

Near the end of the simulations (\rhoxgs{-2}), there is an increase in the mass and momentum of the fast outflows (right-hand column of \figref{fig:outflow}).  For the higher resolution models (\mtsxn{}), a fast, vertical component forms above and below the stellar core (fourth row of \figref{fig:vr}), which ultimately merges with the fast lobes at late times (bottom row of \figref{fig:vr}).  For the lower resolution models (\msxx{}), the increase in mass and momentum is first due to the launching radius of the first core outflow decreasing, and then due to the velocity in the lobes and above/below the core increasing.

In agreement with our previous work \citepeg{WursterLewis2020sc,WursterBateBonnell2021}, stellar core outflows are not launched, at least for as long as we are able to follow the models.

\section{Discussion}
\label{sec:disc}

\subsection{Timescales}
Due to computational limitations, our simulations end 4 to 40~yrs after the formation of the stellar core, which is much shorter than the typical observed age of a young stellar object of \sm$10^6$~yr \citepeg{Bouvier+2014,Ansdell+2016,Froebrich+2018,TeixeiraScholzAlves2020}; even the extremely young stars of \sm$5\times10^3$~yr observed by (e.g.) \citet{FuruyaKitamuraShinnaga2006} and \citet{Yusefzadeh+2017} are still much older than presented here.  Our early end time is limited by our high resolution and densities in the stellar cores, and hence short timesteps (recall \secref{sec:ic:ps}).  Therefore, our conclusions about the origin of magnetic field in stars is based upon a protostar's characteristics at its birth. 

After reaching the maximum magnetic field strength shortly after stellar core formation (\figsref{fig:BVrho}{fig:B:SHC}), there is a rapid decrease in field strength over the first few years; the length of this initial decrease becomes shorter with increasing resolution.  After this rapid decrease, the trend is less clear, given the fluctuations in maximum and stellar core magnetic field strengths.  Thus, predicting how the magnetic field will evolve over the next $10^6$~yr is challenging.

\begin{figure} 
\centering
\includegraphics[width=\columnwidth]{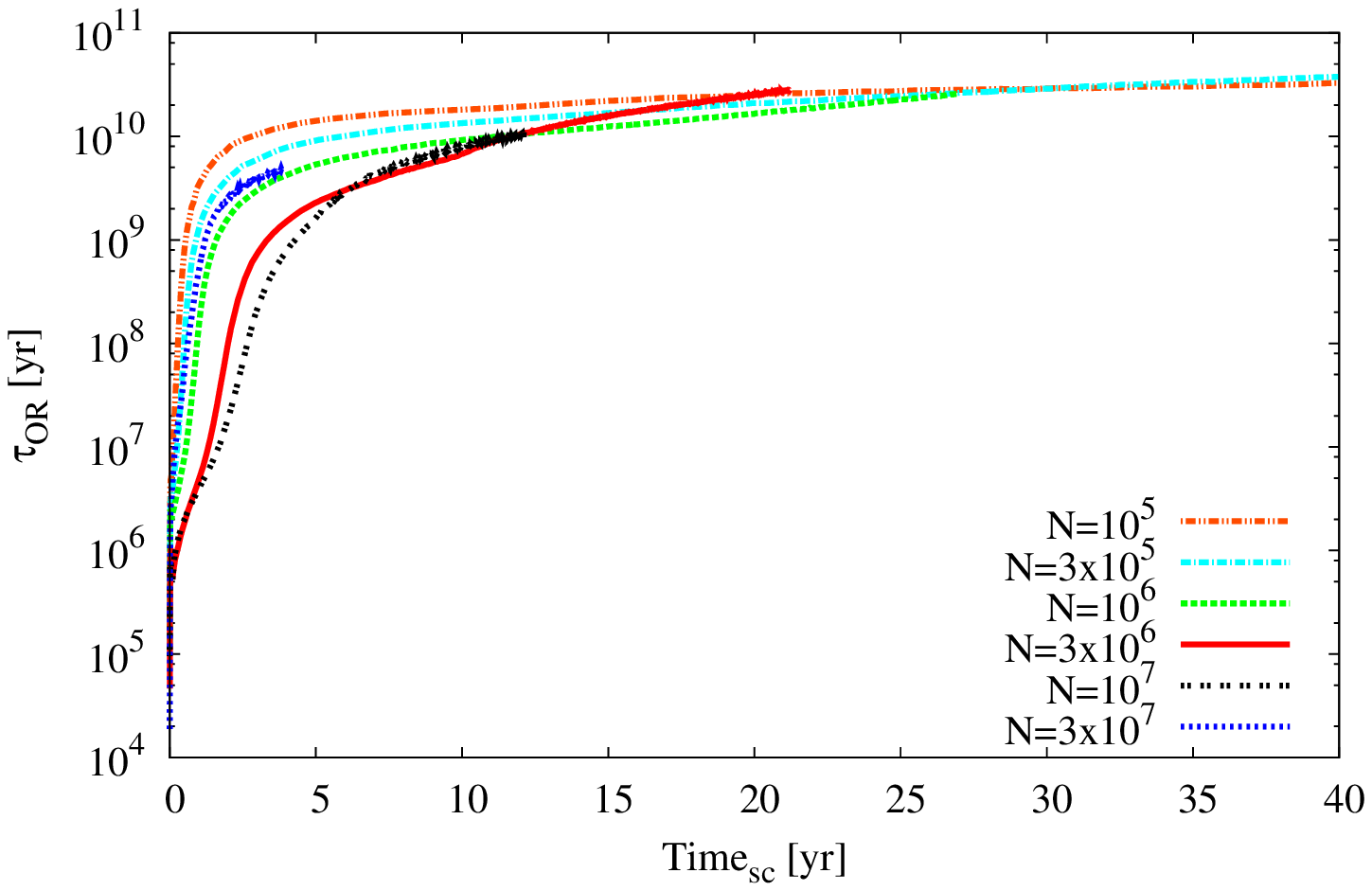}
\caption{The evolution of the Ohmic timescale in the stellar core after stellar core formation, as given by Eqn.~\ref{eq:dtohm}.  The timescale initially increases, before stabilising at $\tau_\text{OR} \sim 10^{10}$~yr.  Since $\tau_\text{OR}  \gg 10^6$~yr, this suggests that the stellar magnetic field will not decay by Ohmic resistivity by the age at which most young stellar objects are observed.  } 
\label{fig:dtOhm}
\end{figure}

At the very high temperatures in and near the stellar core, the gas is highly ionised, therefore Ohmic resistivity is the dominant non-ideal effect.  The diffusion timescale for Ohmic resistivity is given by 
\begin{equation}
\label{eq:dtohm}
\tau_\text{OR} = \frac{L^2}{\eta_\text{OR}},
\end{equation}
where $L$ is the characteristic length scale which we set to the radius of the stellar core, and $\eta_\text{OR}$ is the coefficient for Ohmic resistivity, which is dependent on the number density of the $j$ chemical species that are present $n_j$, their mass $m_j$, their charge $eZ_j$, and their plasma-neutral collisional frequency $\nu_{j\text{n}}$; the coefficient is given by
\begin{equation}
\label{eq:etaohm}
\eta_\text{OR} = \frac{c^2}{4\pi}\left[ \sum_j \frac{n_j \left(eZ_j\right)^2}{m_j \nu_{j\text{n}}} \right]^{-1},
\end{equation}
where $c$ is the speed of light;  see also \citet{WardleNg1999}, \citet{Wardle2007}, \citet{WursterPriceBate2016}, and \citet{Wurster2016}.  At these high temperatures, the present chemical species are likely all gas since the dust grains would have evaporated at lower temperatures \citepeg{LenzuniGailHenning1995}.  Unlike ambipolar diffusion or the Hall effect, this term is independent of the magnetic field strength.  
The evolution of the Ohmic timescale is plotted in \figref{fig:dtOhm}.

In our models, the stellar core radius is $0.01 < r_\text{sc}/\text{au} < 0.015$, therefore the evolution of $\tau_\text{OR}$ is primarily defined by the evolution of Ohmic resistivity. After the initial increase, the Ohmic timescale remains at $\tau_\text{OR} \sim 10^{10}$~yr.  Since $\tau_\text{OR}  \gg 10^6$~yr, the magnetic field in the stellar core is not expected to decrease due to Ohmic resistivity.  Therefore, based upon resistivity arguments, the magnetic field strength in the stellar core shortly after its birth is indicative of its magnetic field strength at later times as well.  This timescale, however, is an upper limit since other processes, such as turbulent diffusion (see \secref{sec:disc:IC:turb}), may contribute to diffusing the magnetic field out of the stellar core.  This would yield a somewhat quicker field decay than just the Ohmic rate, however, it is expected that the strong stellar magnetic field will persist.   

As shown in \figrref{fig:B:SHC}{fig:shc:Bxy:small}, the magnetic field in and around the stellar core is dynamically evolving, with magnetic flux being advected in and out of the stellar core.  Therefore, the magnetic field evolution in the entire region is important to determine the evolution of \Bsc{}.  Since $\eta_{OR}$ increases only slightly outside of the stellar core, assuming a larger radius to encompass the dynamically evolving region would only increase the Ohmic timescale.  Therefore, the fluctuating core field strength is dominated by a combination of magnetic advection and artificial resistivity, and will likely continue to do so until a gap appears between the stellar core and the inner edge of the disc.  Since Ohmic resistivity is un-important on these timescales, the vast majority of the resolution dependence is almost certainly due to the artificial resistivity; this is consistent with higher resolutions yielding stronger magnetic fields.  Thus, our results and the Ohmic diffusion timescale are consistent with strong fields persisting to the ages of observed young stars.

\subsection{Initial conditions}
The results of star formation simulations are inherently dependent on the initial conditions, such as the initial mass, density profile, rotational profile, thermal energy, and magnetic field geometry.  Thus, the stellar field strength may also be dependent on these initial conditions.  We briefly comment on the possible effect of the initial magnetic field strength and turbulence.

\subsubsection{Magnetic field strength}
\label{sec:disc:IC:mag}
Star forming regions are observed to have normalised mass-to-flux ratios of $0.5 \lesssim \mu \lesssim 3$ \citepeg{GirartRaoMarrone2006,Stephens+2013,Koch+2014,Qiu+2014,Hull+2017,Karoly+2020}, although some regions have strengths similar to that presented here \citep[e.g. B335;][]{Maury+2018}.  Cores tend to have larger mass-to-flux ratios than their envelopes \citep{Li+2014}, thus our weaker initial magnetic field strength may better represent a slightly evolved core.  Therefore, how would our conclusions change if we had modelled an initially stronger magnetic field strength?

Naively, one would expect that stronger initial field strengths would lead to stronger maximum and core strengths.  However, the ideal MHD study of \citet{BateTriccoPrice2014} yielded similar maximum magnetic field strengths for both \mueq{5} and 10. 

When non-ideal MHD processes are introduced, predicting magnetic field strengths becomes more challenging.  For example, the effect of ambipolar diffusion depends on the magnetic field strength at low densities, therefore the magnetic dissipation caused by ambipolar diffusion is higher in stronger magnetic fields; moreover the magnetic wall would also be stronger since it is caused primarily by ambipolar diffusion \citepeg{TassisMouschovias2005b,TassisMouschovias2007a,TassisMouschovias2007b,WursterBateBonnell2021}; this may help to prevent the magnetic field from entering the stellar core.  In the cloud-collapse simulations of \citet{Masson+2016} which included ambipolar diffusion, prior to the first core phase, their model with \mueq{2} had a larger distribution of magnetic field strengths at any given density and generally stronger strengths than their model with \mueq{5}.  However, in the first core itself, the magnetic field strengths were slightly higher in the model with \mueq{5}.

Therefore, the effect of increasing the initial magnetic field strength is not straightforward, and additional simulations would be required to test this parameter space (although this is out of the scope of this project).    Based upon the results of \citet{BateTriccoPrice2014} and \citet{Masson+2016}, changing the magnetic field strength will quantitatively change our results, but likely will not qualitatively affect our conclusions.

While predicting the effect of increasing the initial magnetic field strength remains challenging, it is indisputable that including some or all of the non-ideal processes decreases that magnetic field strength compared to ideal MHD models \citepeg{MachidaInutsukaMatsumoto2007,TomidaOkuzumiMachida2015,Tsukamoto+2015oa,WursterBatePrice2018sd,WursterBatePrice2018ff,WursterBateBonnell2021,Masson+2016,MarchandCommerconChabrier2018,Vaytet+2018}, at least for strong initial field strengths.  
How much the field strength is decreased compared to ideal MHD will depend on the microphysics of the non-ideal model \citepeg{Zhao+2020,Zhao+2021,WursterBatePrice2018sd,Wurster2021}.  Nonetheless, non-ideal MHD prevents the unobserved $>100$~kG fields obtained by ideal MHD simulations \citepeg{MachidaInutsukaMatsumoto2006,BateTriccoPrice2014}.  This shows that non-ideal MHD effects are extremely important to reduce the field strength, but are not so dominate as to effectively wipe out a stellar field and seem to give field strengths similar to those that are observed in young stars (with sufficient resolution).

\subsubsection{Turbulent velocity fields}
\label{sec:disc:IC:turb}
Turbulence is typically included in star formation simulations that are initialised from larger cores (e.g. $5 - 10^3$~\Msun{} cores), and has been investigated as a possible solution to the so-called `magnetic braking catastrophe' \citep[i.e. rotationally supported discs do not form in the presence of strong, ideal magnetic fields since well-ordered magnetic fields are very efficient at transporting angular momentum outwards; e.g.][]{AllenLiShu2003,PriceBate2007,MellonLi2008,HennebelleFromang2008}.  Turbulence causes a misalignment between the magnetic field and rotation vectors \citepeg{Joos+2013}, which hinders the outward transport of angular momentum and hence promotes disc formation.  However, turbulence also can act as an effective diffusivity, transporting magnetic flux outwards \citepeg{Santoslima+2012,Santoslima+2013,Joos+2013}.  Thus, turbulence leads to weaker magnetic fields in the first core phase, which may lead to weaker stellar core strengths.

To the contrary, \citet{Seifried+2012,Seifried+2013} argued that turbulence did not act as an effective diffusivity and did not cause a magnetic flux loss, at least on larger scales.  At first core densities and lower, they found $B \propto \rho^{0.5}$, independent of the level of turbulence.  For disc formation, they argued that discs formed simply due to the turbulent velocity structure and not magnetic flux loss.  Unfortunately, this conclusion may have been a result of their use of the mass-to-flux ratio as a diagnostic tool \citep{Santoslima+2013}.  Thus, it would appear that at the first core stage, turbulence likely causes magnetic flux loss and hence lower first core magnetic field strengths.  

The above studies focused on disc formation, and their methods prevented them from investigating the stellar core magnetic field strength.  To the best of our knowledge, \citet{WursterLewis2020sc} is the only study to model the formation of the stellar core from a turbulent magnetised molecular cloud.  This study used the same initial conditions as presented here, but the lower mass resolution of $10^6$ SPH particles in the cloud.  Increasing the initial level of turbulence from Mach 0 to Mach 1 either had negligible effect on the maximum magnetic field strength or decreased it by an order of magnitude, depending on the other initial properties of the sphere and physical processes included.  Since lower \Bmax{} tends to lead to lower \Bfhc{} and \Bsc{} (\figsref{fig:BVrho}{fig:B:SHC}), it is likely that including turbulence will decrease the stellar core magnetic field strength.  

\section{Summary and conclusion}
\label{sec:conc}
In this study, we investigated the effect of numerical resolution on the gravitational collapse of a molecular cloud core through the first and stellar collapse phases to the formation of a protostar.  We tested six resolutions, where each model was initialised as a 1~\Msun{} spherical cloud of uniform density and included between \mf{} and $3\times10^7$ equal-mass SPH particles.  The domain was threaded with an initially vertical magnetic field that was aligned with the rotation axis and had a normalised mass-to-flux ratio of 5.  Our models included Ohmic resistivity, ambipolar diffusion and the Hall effect; the aligned orientation of the magnetic field and rotation vectors meant that the Hall effect would hinder disc formation.

Our main conclusions are as follows:
\begin{enumerate}
\item Increasing resolution increases the length of time it takes to reach stellar densities from the beginning of the simulation, primarily due to a lengthening of the slow, isothermal collapse phase.  The collapse to stellar densities requires similar evolution times at all resolutions when measured from the end of the isothermal collapse at \rhoxeq{-13}.

\item We obtain numerical convergence for the maximum magnetic field strength until late in the first core phase, with the maximum field remaining below 1~G.  The average magnetic field strength in the first core is $\left< B_\text{fhc} \right> \lesssim 10$~G throughout the first core and second collapse phases.

\item At \rhoxapprox{-9}, the maximum magnetic field strength diverges amongst the models due to the formation of a magnetic wall in the outer parts (radii $\approx 3$ au) of the first core.  Following its formation, the maximum field strength resides in the magnetic wall rather than in the centre of the first core.  The wall forms earlier, is better well-defined and less axisymmetric at higher resolutions.  At high numerical resolutions, short whistler waves created by the Hall effect are resolved, which cause the initially axisymmetric wall to become unstable and the magnetic field becomes highly asymmetric. 

\item After the break up of the magnetic wall, the maximum field strength is unconverged, even with our highest resolutions.  By the formation of the stellar core at \rhoxeq{-4}, the maximum field strength differs by a factor of \sm300 between the highest and lowest resolution models (but only a factor of \sm10 between the three highest resolution models).  

\item The magnetic field is dynamically evolving in and around the stellar core, and the field strength fluctuates up to two orders of magnitude after the stellar core's formation.  With increasing resolution, the field strength in the core also increases.  In our highest resolution model, the average magnetic field within the stellar core exceeds 1~kG for 6~months after the formation of the stellar core.  The average magnetic field strengths of the stellar cores in the next two highest resolution models peak around 1~kG, but quickly decay to $\left< B_\text{sc} \right> \lesssim 100$~G.  
In all models, the maximum magnetic field strength lies within a small (radius $\approx 1$~au) disc surrounding the stellar core rather than within the stellar core itself.  For our two highest resolution models, the maximum magnetic field strength exceeds 1~kG for as long as we are able to follow the models.
\item First core outflows are launched in all models with \mtfn{} and are converged for resolutions of  \mtsxn{}. Stellar core outflows do not form in any of our models.
\end{enumerate}

Multiple numerical studies have shown that the initial star forming environment affects the resulting star.  Therefore, our results and hence our conclusions may be affected by (e.g.) turbulence, different initial magnetic field strengths, and magnetic field geometry.  Determining exactly how each of these processes affects the magnetic field implanted at birth would require additional studies.  In our study, the magnetic field strengths during the second collapse phase and within the stellar core increase with increasing resolution.  Numerical convergence of the magnetic field has not been obtained in these phases.  Given the fine structure that develops in the magnetic field due to the Hall effect late in the evolution of the first hydrostatic core, numerical convergence of the magnetic field strength appears to be computational prohibitive to achieve with existing computational resources.  However, given that with our highest resolution calculation we are able to obtain an average magnetic field strength in the stellar core in excess of 1~kG that is sustained for 6 months, we cautiously conclude that substantial magnetic fields may be implanted in low-mass stars during their formation.  Since the Ohmic diffusion timescale in the stellar core is much longer than the age of the young stellar objects that are currently being observed, it is probable that these birth magnetic fields persist over long timescales, suggesting that a dynamo process later in the star's life is not required to generate a strong stellar magnetic field.

\section*{Acknowledgements}

We would like to thank the referee for useful comments that improved the quality of this manuscript.
JW and MRB acknowledge support from the European Research Council under the European Community's Seventh Framework Programme (FP7/2007- 2013 grant agreement no. 339248).  
JW and IAB acknowledge support from the University of St Andrews.
DJP received funding via Australian Research Council grants FT130100034, DP130102078 and DP180104235.  
This work was performed using the DiRAC Data Intensive service at Leicester, operated by the University of Leicester IT Services, which forms part of the STFC DiRAC HPC Facility (www.dirac.ac.uk). The equipment was funded by BEIS capital funding via STFC capital grants ST/K000373/1 and ST/R002363/1 and STFC DiRAC Operations grant ST/R001014/1. DiRAC is part of the National e-Infrastructure.
Several figures were made using \textsc{splash} \citep{Price2007}.  

\section*{Data availability}
The data underlying this article will be available upon reasonable request.
\bibliography{fossil.bib}
\label{lastpage}
\end{document}